\documentclass[onecolumn,fleqn,usenatbib]{mnras}

\usepackage{newtxtext,newtxmath}
\usepackage[T1]{fontenc}

\usepackage{amsmath}
\usepackage{mathtools}
\usepackage{bm}

\numberwithin{equation}{section}

\usepackage{graphicx}

\usepackage[capitalise,compress]{cleveref}
\newcommand{\crefrangeconjunction}{--}
\newcommand{\creflastconjunction}{, and\nobreakspace}
\crefname{section}{Section}{Sections}
\Crefname{section}{Section}{Sections}

\DeclareRobustCommand{\VAN}[3]{#2}
\let\VANthebibliography\thebibliography
\def\thebibliography{\DeclareRobustCommand{\VAN}[3]{##3}\VANthebibliography}

\def\prx{Phys.\ Rev.\ X}

\title[Neutron-star tidal deformability]{{Dynamical neutron-star tides: The signature of a mode resonance}}

\author[P. Pnigouras et al.]{%
    P. Pnigouras$^1$,  N. Andersson$^2$\thanks{N.A.Andersson@soton.ac.uk}, F. Gittins$^{3,4}$ and A. R. Counsell$^2$ \\
        $^1$Departamento de F{\'i}sica Aplicada, Universidad de Alicante, Campus de San Vicente del Raspeig, Alicante 03690, Spain \\
        $^2$Mathematical Sciences and STAG Research Centre, University of Southampton, Southampton SO17 1BJ, United Kingdom \\
        $^3$Institute for Gravitational and Subatomic Physics (GRASP), Utrecht University, Princetonplein 1, 3584 CC Utrecht, Netherlands\\
        $^4$Nikhef, Science Park 105, 1098 XG Amsterdam, Netherlands
}
             
\date{Accepted XXX. Received YYY; in original form ZZZ}

\pubyear{2025}

\begin{document}
\label{firstpage}
\pagerange{\pageref{firstpage}--\pageref{lastpage}}
\maketitle


\begin{abstract}
    Motivated by future opportunities in gravitational-wave astronomy and the ongoing effort to constrain physics under extreme conditions, we consider the signature of individual mode resonances excited during the inspiral of binary systems involving neutron stars. Specifically, we quantify how each resonant mode contributes to the effective (frequency-dependent) tidal deformability. The resonant solution is shown to be accurately represented by a new closed-form approximation, which sheds light on the involved phenomenology, and which should be useful for the development of precise waveform models and future parameter extraction efforts.
\end{abstract}




\section{Introduction} \label{sec:Introduction}

Two neutron stars circle around one another in a cosmic dance that will eventually lead to their mutual demise. As the stars are drawn closer together by the emission of gravitational waves they are increasingly deformed by the tidal interaction. The associated finite-size effects manifest as a dephasing of the gravitational-wave signal, offering valuable information about the internal structure of neutron stars and, in turn, constraining the equation of state of  high density nuclear matter \citep{2008PhRvD..77b1502F,2010PhRvD..81l3016H}. The promise of this was clearly demonstrated with the first direct detection of gravitational waves from a neutron star binary (GW170817; \citealp{2017PhRvL.119p1101A,2018PhRvL.121p1101A}).

The tidal driving---acting on a timescale associated with the orbit---induces the resonant excitation of individual oscillation modes of the star. If the conditions are favourable, such resonances may leave an observable imprint on the gravitational-wave signal from the system. Because of its relatively strong coupling to the tidal potential, the (fundamental) f-mode dominates the star's dynamical tidal response \citep{2016PhRvL.116r1101H,2016PhRvD..94j4028S,2020PhRvD.101h3001A,2021MNRAS.503..533A,2021PhRvR...3c3129S,2024PhRvD.109f4004P,2024PhRvD.109j4064H}, but additional (lower frequency) modes may also come into play. The precise signature of these tidal resonances depends on the detailed, extreme density, neutron-star physics. For example, low-frequency g-modes---owing their existence to internal composition (or entropy) gradients---may be within reach of observations with next-generation gravitational-wave instruments like the Einstein Telescope and Cosmic Explorer \citep{2018PhRvD..97b3016A,2022MNRAS.513.4045K,2023PhRvD.108d3003H,2025MNRAS.536.1967C}. This possibility is exciting because such observations would provide unique insight---beyond global properties like mass and radius---into the nature of matter at extreme densities and pressures. In addition, recent work suggests that a resonance associated with an interface mode (a close relative to g-modes that owes its existence to the presence of a first-order phase transition) may provide a smoking gun signature of the existence of exotic matter---like deconfined quarks---at high densities \citep{2021PhRvD.103f3015L,2024ApJ...964...31M,interface_paper}. However, these additional (low-frequency) resonance features are expected to be faint. Moreover, the properties of each class of oscillation  modes are sensitive to the state and composition of matter, which makes them ideal probes for asteroseismology but also makes the modelling more intricate. This does not mean that we cannot make progress on it. It only means that any predictions should come with caveats.

In this paper we focus on modelling the signature of a resonance and how the individual mode contributions to the star's tidal response can be combined into an ``effective tidal deformability'' measure. The idea is simple. We know that the neutron star tide has two main components; a static contribution \citep[represented by the familiar Love numbers;][]{2008ApJ...677.1216H,2008PhRvD..77b1502F} and a dynamical contribution (corresponding to mode resonances and dominated by the star's fundamental oscillation mode; see references above). The pre-resonance evolution is dominated by the equilibrium tide, with the star's response locked to the tidal driving. As the system approaches a resonance, the free oscillations of the star are excited, impacting on the post-resonance solution. The resonant excitation of the star's oscillation modes by the time-dependent tidal driving during the binary inspiral leads to a complicated ``response function'' which encodes the physics of the neutron star. The argument is firmly established---see, for instance, \citet{1994MNRAS.270..611L}, \citet{1994ApJ...426..688R} and \citet{1995MNRAS.275..301K}---especially in the context of Newtonian gravity, and the phenomenology is fairly well understood. In this paper we promote the view that the total tidal imprint should be considered in terms of a time/frequency-dependent effective tidal deformability and establish for the first time the precise signature of each mode resonance as a closed-form, analytical expression. Making use of well-justified assumptions and approximations, this expression works well for all relevant modes, avoiding divergent behaviour near the resonances, and may be used to combine the dynamical contributions of all the modes into a total effective tidal deformability measure. This should, in turn, help inform the development of the robust gravitational-waveform models required to establish the detectability of individual resonances and, perhaps in the extension, facilitate sensitive searches for these features in observational data. This problem is expected to become relevant in the next-generation era, with instruments like the Einstein Telescope \citep{2025arXiv250312263A} and Cosmic Explorer \citep{2019BAAS...51g..35R} able to eke out fine-print details of binary neutron-star inspiral signals.

The numerical solution of the problem is straightforward but of limited use if we want to explore the actual dependence on the parameters of the problem for gravitational-waveform modelling and parameter estimation. We are thus led to consider approximate, analytic solutions. Previous work in this direction explored the pre- and post-resonance behaviour \citep{1994MNRAS.270..611L} as well as solutions valid through resonance obtained via matched asymptotics (see, in particular, the steps laid out by \citealt{2005MNRAS.357..834R}, \citealt{2007PhRvD..75d4001F} and \citealt{2020PhRvD.101j4028P}). For our purposes it makes sense to highlight the work by, first of all, \citet{2016PhRvL.116r1101H} and \citet{2016PhRvD..94j4028S}, and, secondly, \citet{2024PhRvD.110b4039Y}. The former studies follow a strategy very similar to the one we adopt in the following \citep[also notable for the fact that the result has been implemented in a state-of-the-art waveform model for gravitational-wave analysis;][]{2024PhRvD.109b4062A}. The latter effort involves a slightly different approach based on isolating the equilibrium tide from the mode excitation (involving a clever resummation of the solution) and sheds light on the post-resonance features. The post-resonance behaviour---not explored by \citet{2016PhRvL.116r1101H} and \citet{2016PhRvD..94j4028S}, as their main focus was the on the fundamental mode which is unlikely to become resonant during binary inspiral---is particularly important for low-frequency mode resonances. Our aim here is to explore all relevant aspects in detail. The analysis will lead us to a closed-form expression offering considerable intuitive insight and which is demonstrated to be accurate to a few percent compared to the numerical solution of the problem. Like previous work using the mode-sum formalism for the description of the dynamical tide \citep{2020PhRvD.101h3001A,2024MNRAS.527.8409P}, the basic arguments are not tailored to particular assumptions about the neutron star physics and can thus be easily extended to include effects relating to, e.g., the neutron star's elastic crust \citep{2012PhRvL.108a1102T, 2021MNRAS.504.1129N,2021MNRAS.504.1273P}, superfluid components \citep{2017MNRAS.464.2622Y,2022MNRAS.514.1494P}, exotic matter \citep{2017MNRAS.464.2622Y}, or even relating to other types of systems, e.g. white dwarfs \citep{2011MNRAS.412.1331F,2012MNRAS.421..426F}. Furthermore, given that it allows us to easily track the tidal response as the system evolves through resonance, our results provide a natural foundation for future exploration of the impact on the gravitational-wave signal.


\section{Formulating the tidal problem}


\subsection{The mode-sum approach} \label{sec:The mode-sum approach to the tidal problem}

In Newtonian gravity, the tidal response of the (non-rotating) primary star due to the presence of its binary companion is obtained from a solution to the linear perturbation equations \citep[see, e.g.,][]{1994MNRAS.270..611L}:
\begin{gather}
    \partial_t^2 \xi_i + \frac{1}{\rho} \nabla_i \delta p - \frac{1}{\rho^2} \delta \rho \nabla_i p  + \nabla_i \delta \Phi = - \nabla_i \chi, \label{Euler perturbed} \\
    \delta \rho + \nabla_i (\rho \xi^i) = 0, \label{continuity perturbed} \\
    \nabla^2 \delta \Phi - 4 \pi G \delta \rho = 0, \label{Poisson perturbed}
\end{gather}
where $\xi^i$ is the Lagrangian displacement vector that characterises the motion of the fluid elements, and $\rho$ and $p$ are  the mass density and pressure of the equilibrium star, respectively. The tidally-induced (Eulerian) fluid perturbations are indicated by $\delta$  and $\delta \Phi$ represents the perturbed gravitational potential, with $G$ the gravitational constant.
The system of fluid equations is closed by a thermodynamic relation that encodes the microphysics of the matter; the equation of state.

The tidal potential, $\chi$, is naturally expanded in spherical harmonics associated with  a spherical coordinate system $(r, \theta, \phi)$ centred on the primary star and with its $z$ axis aligned with the orbital angular momentum \cite[see, e.g.,][]{1977ApJ...213..183P}
\begin{equation}
    \chi=  - \frac{G M'}{|x^i - D^i(t)|} =-\sum_{l\geq 2}\sum_{m=-l}^l v_{lm} r^l Y_{lm} e^{-im\Phi(t)} \ , 
    \label{tidpot}
\end{equation}
where
\begin{equation}
    v_{lm} =  \frac{GM' W_{lm}}{D^{l+1}} \ ,
\end{equation}
and $M'$ is the mass of the (here point-like) binary companion, $D(t)$ is the orbital separation, $\Phi(t)$ is the orbital phase (not to be confused with the gravitational potential!), and $Y_l^m$ are spherical harmonics of multipolar degree $l$ and azimuthal order $m$. The constant $W_{l m}$ is non-zero only for even $l+m$ and is given by
\begin{equation}
    W_{l m} 
        = (- 1)^{(l + m) / 2} \sqrt{\frac{4 \pi}{2 l + 1} (l - m)! (l + m)!} 
            \left[ 2^l \left( \frac{l + m}{2} \right)! 
            \left( \frac{l - m}{2} \right)! \right]^{- 1}.
    \label{W_lm}
\end{equation}
For the leading order contribution to the gravitational-wave signal ($l=2$) we need the coefficients $W_{20}=-\sqrt{\pi/5}$ and $W_{2\pm 2}=\sqrt{3\pi/10}$, while $W_{2\pm 1}=0$.

In the mode-sum approach to the tidal problem \citep{1994MNRAS.270..611L,1995MNRAS.275..301K,2020PhRvD.101h3001A,2024MNRAS.527.8409P}, the solution to \cref{Euler perturbed,continuity perturbed,Poisson perturbed} is expressed in terms of the free oscillation modes of the star. This is natural since, at least in the absence of dissipation, the modes form a complete basis (see, e.g., \citealp{1978ApJ...221..937F} and \citealp{2001PhRvD..65b4001S}). The solution is then expressed as a sum over all normal modes (each labelled by $n$)
\begin{equation}
    \xi^i(t, x^i) = \sum_n a_n(t) \xi_n^i(x^i).
    \label{mode sum}
\end{equation}
For non-rotating stars, the mode solutions $\{ \xi_n \}$ are complex solutions to \cref{Euler perturbed} corresponding to (real) eigenvalues $\{ \omega_n^2 \}$. The fact that the eigenvalue is $\omega_n^2$, rather than $\omega_n$, is important because it leads to the mode solutions of a non-rotating star being independent of the sign of the mode frequency. Moreover, given the $e^{im\varphi}$ azimuthal dependence associated with the spherical harmonics, it is easy to show that a (real-frequency) mode solution and its complex conjugate are orthogonal (as they must be in order for the mode sum to be able to represent a real-valued function). In order to obtain a complete basis for the tidal response we only need to include solutions with positive frequencies, $\omega_n \ge0$, but we have to account for both of the $\pm m$ solutions in the mode sum. However, the negative $m$ solutions are easily included as the complex conjugates of the positive $m$ ones. This is the convention we adopt in the following. Moreover, we label each oscillation mode by $n$, implicitly including the $l$ and $m$ of the relevant spherical harmonic (only making these labels explicit whenever it helps understanding). That is, we generally use $\omega_n$ rather than $\omega_{nlm}$ (and similar for other variables associated with the modes). This convention makes all expressions more compact and should not cause any confusion. 

Effectively, the mode sum transforms the tidal problem into that of a driven harmonic oscillator (with a time-varying driving frequency), where each mode amplitude, $a_n$, is governed by
\begin{equation}
	\ddot{a}_n(t)+\omega_n^2 a_n(t)=\frac{v_{lm}(t) Q_{n}}{\mathcal A_n^2}e^{-im\Phi(t)} .\label{mode equation of motion}
\end{equation}
The equation of motion~\eqref{mode equation of motion} is obtained by substituting \cref{mode sum} into \cref{Euler perturbed} and using the mode eigenvalue equation (see \citealt{2001PhRvD..65b4001S} or \citealt{2024MNRAS.527.8409P} for the detailed steps).

Each mode's contribution to the orbital evolution is encoded in the mass multipole moment associated with the fluid perturbation:
\begin{equation}
    Q_{n} =\int_0^R \delta \rho_n  r^{l + 2} dr,
\end{equation}
where $\delta \rho_n (r)$ encodes the radial dependence of the  density perturbation associated with a specific mode, obtained after expanding perturbations as $\delta\rho_n(x^i)=\delta\rho_n(r)Y_l^m(\theta,\phi)$ (and likewise for other quantities). For real-frequency modes, $\delta\rho_n(r)$ and hence $Q_{n}$ are both real. Moreover, the constant
\begin{equation}
    \mathcal A_n^2=\int \xi_n^{i *} \xi_{n \, i} \rho dV \label{mode normalisation}
\end{equation}
represents the chosen mode normalisation. Finally, each individual fluid mode is associated with an energy
\begin{equation}
 E_n  = {1\over 2} \left( |\dot a_n|^2 +\omega_n^2 |a_n|^2 \right) \mathcal A_n^2.
 \label{modEn}
\end{equation}

The formulation of the problem is standard and the nature of the associated solutions is well understood, both via approximations and numerical solutions. It is nevertheless worth revisiting this territory given the potential detectability of individual resonances with next-generation gravitational-wave instruments. This is an important question given that features associated with g-modes (for example) may provide insight into the state and composition of matter at supranuclear densities. So far, the discussion  has been based on relatively simple energy estimates \citep{2018PhRvD..97b3016A,2022MNRAS.513.4045K,2023PhRvD.108d3003H,2025MNRAS.536.1967C}. In order to draw more robust conclusions, we need to to better. Our aim here is to develop a new approximate solution for the mode amplitude as the system evolves through a given resonance, benchmark this solution against  numerical results and hence develop a model for the contribution that each mode makes to the effective Love number/tidal deformability. Provided this solution is simple enough---and we will argue that it is!---we would then have a useful foundation for future work on the corresponding gravitational-waveform models.

At this point it is worth noting that, as shown by \citet{2001PhRvD..65b4001S}, the mode sum \eqref{mode sum} is not practically useful in the case of rotating stars, because the mode equations of motion \eqref{mode equation of motion} couple to each other, owing to the fact that the modes obey a modified orthogonality condition. To avoid this, the displacement and its time derivative should be expanded simultaneously (in a phase space decomposition). For completeness, we present the relevant arguments in \cref{sec:phase space}, where we express the Love number and the energy in phase space, obtaining the corresponding relations for rotating stars. The derivation of the relevant expressions in phase space is also necessary for comparing the results obtained here to those by \citet{2024PhRvD.110b4039Y}, a point we return to later.


\subsection{The equilibrium tide} \label{sec:The equilibrium tide}

It is useful to start by considering the orbital evolution. In the absence of tides, we know from the (leading-order) quadrupole formula for gravitational-wave emission that the binary orbit evolves according to
\begin{equation}
    \dot \Omega = {96\over 5c^5} (G\mathcal M)^{5/3} \Omega^{11/3} \ , 
    \label{radrec}
\end{equation}
where the orbital frequency $\Omega$ follows from Kepler's law
\begin{equation}
    \Omega^2 = {GM\over D^3} \ . 
\end{equation}
Here $M=M_\star+M'$ is the total mass of the system,
\begin{equation}
    \mathcal M = {(M_\star M')^{3/5} \over M^{1/5}} \ ,
\end{equation}
is the so-called chirp mass, while $M_\star$ is the mass of the primary. \Cref{radrec} reminds us of the well-known fact that the leading-order post-Newtonian gravitational-wave signal only allows us to determine this specific combination of the individual masses. If we want to go beyond this, then we need to consider higher-order contributions to the signal. 

We see that we may introduce a characteristic inspiral timescale
\begin{equation}
    t_{D} = \left( {\Omega \over \dot \Omega}\right)_{\Omega=\Omega_0} = {5c^5 \over 96} \left( G \mathcal M\right)^{-5/3} \Omega_0^{-8/3} \ ,
    \label{timeD}
\end{equation}
with $\Omega_0$ the initial orbital frequency. Putting typical numbers in, this leads to 
\begin{equation}
    t_{D} \approx 0.9 \left( {1.2M_\odot\over \mathcal M}\right)^{5/3} \left( {100\ \mathrm{Hz} \over \Omega_0/2\pi}\right)^{8/3}\ \mathrm{s} \ .
\end{equation}
We have scaled the result to the chirp mass expected for an equal mass $1.4M_\odot$ binary \citep[the value is also close to that for GW170817;][]{2017PhRvL.119p1101A,2018PhRvL.121p1101A,2019PhRvX...9a1001A} and a frequency expected of the low-order g-modes of a realistic neutron star \citep[see, e.g.,][]{2025MNRAS.536.1967C}. This estimate suggests that these modes would become resonant about a second before merger, i.e. very late in the binary inspiral.

It is also easy to argue that we may (safely) treat the inspiral as adiabatic. Specifically, Taylor expanding the orbital separation around $t_0$, some chosen time when the binary evolution begins, we have
\begin{equation}
    D(t) = D(t_0) + \dot D(t_0) (t-t_0) + ... \approx  D_0 \left[ 1 + {\dot D (t_0)\over D(t_0)} (t-t_0) \right]
     \approx D_0 \left[ 1 - {2\over 3} \left({\dot \Omega \over \Omega} \right)_{\Omega_0} (t-t_0) \right] = D_0 \left( 1 - {2\over 3} {t-t_0\over t_{D}} \right) \ .
\end{equation}
That is, as long as $t-t_0 \ll t_{D}$ we have
\begin{equation}
    {1\over D^{l+1}(t)} \approx {1\over D_0^{l+1}} \left[ 1 + {2(l+1)\over 3} {t-t_0\over t_{D}} \right] \ .
\end{equation}
Effectively, we may take this to be constant (to leading order). Then, \cref{mode equation of motion} becomes
\begin{equation}
	\ddot{a}_n(t)+\omega_n^2 a_n(t) \approx \frac{v_{lm}(t_0) Q_{n}}{\mathcal A_n^2}e^{-im\Phi(t)} \ .
\label{driven}
\end{equation}
In essence, the time dependence of the tidal driving---which must be inherited by the mode solution, at least before resonance---is entirely associated with the phase $\Phi(t)$. Moreover, given that $\dot \Phi = \Omega$, it follows immediately that the (particular) solution to the problem is
\begin{equation}
   a_n(t) \approx  \frac{Q_{n}v_{lm} (t_0)}{\mathcal A_n^2}\frac{e^{-im\Phi(t)}}{\omega_n^2-(m\Omega)^2} \ .
   \label{equiltide}
\end{equation}
This solution represents the instantaneous response to the tidal driving, often referred to as the equilibrium tide. The corresponding mode energy is
\begin{equation}
    E_n \approx {1\over 2}  \frac{Q_{n}^2}{\mathcal A_n^2}  \left( {GM' W_{lm}\over D^{l+1}}\right)^2
    { \omega_n^2 + m^2\Omega^2  \over (\omega_n^2 - m^2 \Omega^2)^2}  \ .
    \label{pre_energy}
\end{equation}
The static contribution is obtained by setting $\Omega=0$ in this expression. 

In gravitational-wave astronomy, the tidal response is commonly quantified in terms of the tidal deformability, which derives from the so-called Love number. Within the mode-sum approach, it is easy to write down the expression for the effective Love number. Starting from the multipole expansion of the tidal potential \eqref{tidpot},
\begin{equation}
    \chi = \sum_{lm} \chi_{lm} Y_l^m \ , 
\end{equation}
the Love number, $k_{l m}$, quantifies the extent to which the star is deformed by the tidal potential. It is defined by the relation 
\begin{equation}
    \delta \Phi_{l m} = 2 k_{l m} \chi_{l m},
    \label{efflove}
\end{equation}
where $\delta\Phi_{l m}$ is the gravitational potential perturbation induced by $\chi_{l m}$, both evaluated at the stellar surface. For a time-varying tidal field the Love number encodes the dependence on the orbital motion and  can be used to represent both the static and the dynamical tide.

Noting that
\begin{equation}
    \delta \Phi = \sum_n a_n \delta \Phi_n
\end{equation}
and using the fact that
\begin{equation}
    \delta \Phi_n(R) = - \frac{4 \pi G}{(2 l + 1) R^{l + 1}} Q_n
    \label{eq:GravitationalPotentialMultipoleMomentRelation}
\end{equation}
(which follows after using the Green's function for Poisson's equation \eqref{Poisson perturbed}) we get
\begin{equation}
    \delta \Phi = 
    -  \sum_n \frac{4 \pi G}{(2 l + 1) R^{l + 1}}  a_n Q_n Y_l^m
        = - 4 \pi G \sum_n \frac{1}{(2 l + 1) R^{l + 1}} 
            \frac{ v_{lm} Q_n^2}{\mathcal{A}_n^2  
            [\omega_n^2 - (m \Omega)^2]} 
            e^{- i m \Phi(t)} Y_l^m\ .
\end{equation}
The sum in this expression includes all the different oscillation modes of the star. Labelling the subset of modes (f,\ p,\ g, ...) for a given harmonic $(l,m)$ by $n'$ (a convention we adopt from here on, see \citealt{2024MNRAS.527.8409P} for additional discussion) we have
\begin{equation}
    \delta \Phi_{l m}(t, R)
        = - \frac{4 \pi G}{(2 l + 1) R^{l + 1}} v_{l m} 
            e^{- i m \Phi(t)} 
            \sum_{n'} \frac{Q_{n'}^2}{\mathcal{A}_{n'}^2 
            [\omega_{n'}^2 - (m \Omega)^2]} \ , 
\end{equation}
and it follows from \cref{efflove} that the mode-sum expression for the tidal response is 
\begin{equation}
    k_{lm} =
    - {1\over 2 v_{lm} e^{-im \Phi(t)} R^l}\delta \Phi_{lm}
        = \frac{2 \pi G}{(2 l + 1) R^{2 l + 1}} 
            \sum_{n'} \frac{Q_{n'}^2}{\mathcal{A}_{n'}^2 
            [\omega_{n'}^2 - (m \Omega)^2]}.
    \label{eq:Love}
\end{equation}

It is convenient to express the result in terms of dimensionless quantities (identified with a tilde here and in the following). To effect this, we  use the convention that
\begin{equation}
    \mathcal A_n^2 = M_\star R^2,
\end{equation}
where  $R$ is the radius of the primary star, and
introduce
\begin{equation}
    Q_n = M_\star R^l \tilde Q_n,
\end{equation}
as well as
\begin{equation}
    \omega_n^2 = \left({GM_\star \over R^3} \right) \tilde \omega_n^2
\end{equation}
(and similar for $\tilde \Omega^2$, although we note that the stellar density is obviously not a natural parameter to use for the normalisation of the orbital frequency).We then have
\begin{equation}
    k_{l m} =  \frac{2 \pi }{2 l + 1 }
            \sum_{n'} {\tilde Q_{n'}^2  \over 
            \tilde \omega_{n'}^2 - (m \tilde \Omega)^2} , 
    \label{kLove}
\end{equation}
representing the effective (manifestly frequency-dependent and dimensionless) Love number for the equilibrium tide. The solution is evidently valid only away from resonance. If we want an expression valid through resonance we need to do more work.

The static Love number follows in the limit $\Omega\to 0$ and the contribution from each mode is simply given by (noting the degeneracy in $m$)
\begin{equation}
    k_{n}^0 =  \frac{2 \pi G}{(2 l + 1) R^{2 l + 1}} \frac{Q_{n}^2}{\mathcal A_{n}^2 \omega_{n}^2} 
        = \frac{2 \pi }{2 l + 1} 
            \frac{\tilde Q_{n}^2}{
            \tilde \omega_{n}^2}. 
    \label{StaticLove}
\end{equation}

Before we proceed it is useful to comment on the validity of \cref{equiltide}. Introducing a new amplitude variable (also helpful for the numerical solutions)
\begin{equation}
    a_n = b_n e^{-im\Phi} \ ,
\end{equation}
we see that \cref{driven} is replaced by
\begin{equation}
    \ddot b_n - 2im\Omega \dot b_n + \left( \omega_n^2 - m^2 \Omega^2-im\dot \Omega \right) b_n = \frac{v_{lm}(t_0) Q_{n}}{\mathcal A_n^2} \ , 
\end{equation}
and it follows that the equilibrium-tide approximation requires, in particular,
\begin{equation}
    |\omega_n^2 - m^2 \Omega^2| \gg |m\dot \Omega| \ .
\end{equation}
This condition clearly cannot be satisfied close to resonance (as the left-hand side then vanishes identically). In order to bridge the evolution across the resonance, we need a different approximation.


\section{Approximate analytical solution} \label{sec:Approximate analytical solution}

In principle, it is straightforward to solve the equation of motion~\eqref{driven}. The problem simply represents a driven harmonic oscillator (although, with a time-varying driving frequency) and the  general solution takes the form $a_n=a_n^\mathrm{(H)}+a_n^\mathrm{(P)}$, where
\begin{equation}
	a_n^\mathrm{(H)}=\tilde a_n^{(1)}e^{i\omega_n t}+\tilde a_n^{(2)}e^{-i\omega_n t} \label{homsol}
\end{equation}
is the solution to the homogeneous problem (representing the free oscillations of the body) and
\begin{equation}
	a_n^\mathrm{(P)}=-\frac{ie^{i\omega_n t}}{2\omega_n}\frac{Q_{n}}{\mathcal A_n^2}\int_{t_0}^t v_{lm} e^{-i(\omega_n t'+m\Phi)} d t'+\frac{ie^{-i\omega_n t}}{2\omega_n}\frac{Q_{n}}{\mathcal A_n^2}\int_{t_0}^t v_{lm} e^{i(\omega_n t'-m\Phi)}d t' \label{partsol}
\end{equation}
is a particular solution obtained via the method of the variation of parameters (with $t_0$ some suitable reference time). Taking this as our starting point, we want to develop an approximate solution for the mode amplitudes that remains valid as the system evolves through an individual resonance.

The first step is (naturally) provided by the equilibrium tide from \cref{equiltide}. If there is no resonance in the time interval of interest, then the  general solution to \cref{driven} takes the form
\begin{equation}
	a_n = \tilde{a}_n^{(1)} e^{i\omega_n t} + \tilde{a}_n^{(2)} e^{-i\omega_n t}
        + \frac{Q_{n}v_{lm} (t)}{\mathcal A_n^2}\frac{e^{-im\Phi}}{\omega_n^2-(m\Omega)^2}.
    \label{before}
\end{equation}
Evidently, we need  the (constant) amplitudes of $e^{\pm i\omega_n t}$ to vanish in the regime where \cref{equiltide} applies. This simply means that the individual modes are not excited before resonance, which makes sense because the frequency of the tidal driving is not (yet) well matched to that of the mode we are focussing on.


\subsection{Near resonance solution} \label{subsec:near_resonance_solution}

In order to develop a solution valid through resonance, let us focus on positive mode frequencies and $m\ge 0$, which means that the resonance condition becomes $\Omega \approx \omega_n/m$. Using the stationary phase approximation (see \citealt{2016PhRvD..94j4028S} for a similar calculation) to evaluate the particular solution \eqref{partsol} we see that the integral that is unaffected by the resonance---with our conventions, the first term---oscillates rapidly and hence averages to zero, so it can be neglected. The other integral also oscillates rapidly, except in the vicinity of the resonance, which is where the main contribution is made.

With this behaviour in mind, we Taylor expand around the resonance to get
\begin{equation}
    \Omega(t)\approx \frac{\omega_n}{m}+\dot{\Omega}(t_n)(t-t_n), \label{Omega expansion near resonance}
\end{equation}
where $t_n$ is the time at resonance. We then have
\begin{equation}
    - 2m \omega_n \dot{\Omega}(t_n) (t-t_n) \approx \omega_n^2 - m^2 \Omega^2 \ .
    \label{Omndot}
\end{equation}
The orbital phase is then given by (note that, as we expand with respect to $t=t_n$, this has a different reference phase compared to the equilibrium tide solution from before)
\begin{equation}
    \Phi(t) = \Phi(t_n)+\int_{t_n}^t\Omega(t') dt'
        \approx \Phi(t_n)+\frac{\omega_n}{m}(t-t_n)+\dot{\Omega}(t_n)\frac{(t-t_n)^2}{2}.
    \label{nearphase}
\end{equation}

For the case we are considering we need
\begin{equation}
     i(\omega_n t - m\Phi) \approx  - i m \Phi(t_n)  + i \omega_n t_n  - im \dot{\Omega}(t_n)\frac{(t-t_n)^2}{2} ,
\end{equation}
which means that
\begin{equation}
    \int_{t_0}^t v_{lm} e^{ i(\omega_n t' - m\Phi)} dt' \approx
	v_{lm}(t_n)e^{-im\Phi_n}\int_{t_0}^t e^{-i(m/2)\dot{\Omega}(t_n)(t'-t_n)^2} dt',
\end{equation}
where we have defined
\begin{equation}
    \Phi_n= \Phi(t_n) - \frac{\omega_n}{m} t_n.
\end{equation}
Recall that we assume that $v_{lm}$ varies slowly near the resonance, in order to treat it as a constant in the integral. This argument should be valid as long as the orbital evolution remains adiabatic near resonance. This is expected to be the case for all low-frequency resonances.

Making use of the the near-resonance expansion the solution can now be written
\begin{equation}
	a_n = \left[
     {i \over 2 \omega_n} \frac{Q_n}{\mathcal A_n^2} v_{lm}(t_n) e^{i(m/2)\dot{\Omega}(t_n)(t-t_n)^2}\int_{t_0}^t e^{-i(m/2)\dot{\Omega}(t_n)(t'-t_n)^2} dt' \right] e^{-im\Phi(t)}.
\end{equation}
Introducing $u=t-t_n$ and assuming that the resonance occurs a long time after the evolution is initiated (at $t=t_0$), we can set $(t_0-t_n)\to-\infty$ to get 
\begin{equation}
	a_n = \left[
     {i \over 2 \omega_n} \frac{Q_n}{\mathcal A_n^2} v_{lm}(t_n) e^{i(m/2)\dot{\Omega}(t_n)u^2}\int_{-\infty}^u e^{-i(m/2)\dot{\Omega}(t_n){u'}^2 }du' \right] e^{-im\Phi(t)},
     \label{nearsol}
\end{equation}
where the integral can be expressed in terms of the Gauss error function:
\begin{equation}
	\int_{-\infty}^u e^{-i(m/2)\dot{\Omega}(t_n){u'}^2} du' = \sqrt{\frac{\pi}{2im\dot{\Omega}(t_n)}}\left[1+\mathrm{erf}\left(\sqrt{\frac{im\dot{\Omega}(t_n)}{2}}u\right)\right].
    \label{mode amplitude near-resonance solution integral}
\end{equation}
Noting the asymptotic expansion (for $w\to\infty$)
\begin{equation}
    \mathrm{erf} (w) \approx 1 - {e^{-w^2} \over \sqrt{\pi} w},
\end{equation}
we have, well before resonance (when $u$ is negative)
\begin{equation}
	\int_{-\infty}^{u} e^{-i(m/2)\dot{\Omega}(t_n){u'}^2} du' 
    = \sqrt{\frac{\pi}{2im\dot{\Omega}(t_n)}}\left[1-\mathrm{erf}\left(\sqrt{\frac{im\dot{\Omega}(t_n)}{2}}|u|\right)\right],
\end{equation}
which leads to (for large  $|u|$)
\begin{equation}
	\int_{-\infty}^{u} e^{-i(m/2)\dot{\Omega}(t_n){u'}^2} du' \approx    \frac{e^{-i(m/2)\dot{\Omega}(t_n){u}^2}}{im\dot{\Omega}(t_n)|u|}.
    \label{Pre}
\end{equation}
This is an important result because it allows us to make the required connection with the equilibrium tide. By comparing to \cref{nearsol} we note that the exponential factors cancel in the asymptotic regime. If we simply undo the near-resonance Taylor expansion, the solution reduces to \cref{equiltide}. Specifically, we have
\begin{equation}
	a_n \approx 
      - {1 \over 2 \omega_n} \frac{Q_n}{\mathcal A_n^2} v_{lm}(t_n)  {1\over m\dot \Omega(t_n) u}   e^{-im\Phi(t)} ,
\end{equation}
leading to, once we use \cref{Omndot} and replace $v_{lm}(t_n)$ with $v_{lm}(t)$,
\begin{equation}
	a_n
    \approx   \frac{Q_n}{\mathcal A_n^2} v_{lm}(t) {  e^{-im\Phi(t)} \over \omega_n^2 - m^2\Omega^2}.
\end{equation}


\subsection{Beyond resonance} \label{subsec:beyond_resonance}

In order to capture the post-resonance behaviour of the mode amplitude solution, we may follow the steps from the near-resonance case, except that we no longer replace the orbital phase (outside the integral) with the Taylor expansion \eqref{nearphase}. Instead, we use (as before, for $m>0$)
\begin{equation}
	a_n = 
     {i \over 2 \omega_n} \frac{Q_n}{\mathcal A_n^2} v_{lm}(t_n) e^{-i(\omega_n t + m\Phi_n)}\int_{-\infty}^u e^{-i(m/2)\dot{\Omega}(t_n){u'}^2} du'
\end{equation}
(noting that, for the evaluation of the integral the Taylor expansion of the phase still holds, as we are using the stationary phase approximation).

In the limit when $u$ becomes large the integral becomes
\begin{equation}
   \int_{-\infty}^u e^{-i(m/2)\dot{\Omega}(t_n){u'}^2} du' \approx \left[\frac{2\pi}{im\dot{\Omega}(t_n)} \right]^{1/2} - \frac{e^{-i(m/2)\dot{\Omega}(t_n){u}^2}}{im\dot{\Omega}(t_n)u} . \label{Post}
\end{equation}
The first term leads to
\begin{equation}
    a_n =  {i\over 2\omega_n} \frac{Q_n}{\mathcal A_n^2} v_{lm}(t_n) \left[ \frac{2\pi}{m\dot{\Omega}(t_n)} \right]^{1/2} e^{- i(\omega_n t + m\Phi_n + \pi/4)} = \tilde a_n e^{-i\omega_n t}.
    \label{mode amplitude post-resonance solution at infinity}
\end{equation}
We see that the resonantly excited mode now oscillates at its natural frequency (at a constant amplitude).

The second contribution to the solution  leads to the equilibrium tide (the argument for which should still hold away from the resonance). To demonstrate this, we need 
\begin{equation}
	a_n = \tilde a_n e^{-i\omega_n t} -  
     {1 \over 2 \omega_n} \frac{Q_n}{\mathcal A_n^2} v_{lm}(t_n) { e^{-im\Phi(t)} \over m\dot{\Omega}(t_n)u}.
\end{equation}
We then use the arguments from the pre-resonance solution (undo the Taylor expansion and replace $v_{lm}(t_n)$ by $v_{lm}(t)$) to arrive at 
\begin{equation}
 a_n = \tilde a_n e^{-i\omega_n t} +  
     \frac{Q_n v_{lm}(t)}{\mathcal A_n^2}  { e^{-im\Phi} \over\omega_n^2 -m^2 \Omega^2}.
\end{equation}
We now recognize the second term as the equilibrium tide.


\subsection{A single, closed-form expression}

The main take-home message at this point is that we can, by paying careful attention to the asymptotic behaviour, extend the near-resonance solution \eqref{nearsol} into the asymptotic regime (both before and after resonance) in such a way that we make contact with the expected form of the solution. This is entirely in the spirit of matched asymptotics expansions and it is crucial as it allows us to write down a single, closed-form expression for the tidal response associated with the mode resonance. However, in order to make the result  more ``practically'' useful it may be better to work with real-valued functions and build the model in terms of Fresnel integrals (adopting the strategy used by \citealp{2005MNRAS.357..834R}, \citealp{2007PhRvD..75d4001F} and \citealp{2016PhRvD..94j4028S}). The expressions involving the error function were helpful in identifying the equilibrium tide contributions to the solution, but now that the strategy is laid down we can easily repeat the  steps for the Fresnel integrals.

As the detailed calculation involves somewhat lengthy expressions we have relegated it to \cref{sec:derivation}. The steps involved are, however, fairly obvious. First, we rewrite the integral we need in \cref{nearsol} as
\begin{equation}
	\int_{-\infty}^u e^{-i(m/2)\dot{\Omega}(t_n){u'}^2} du' = \sqrt{\frac{\pi}{m\dot{\Omega}(t_n)}}\left\{\frac{1}{2}+\mathcal{C}\left(\sqrt{\frac{m\dot{\Omega}(t_n)}{\pi}}u\right)-i\left[\frac{1}{2}+\mathcal{S}\left(\sqrt{\frac{m\dot{\Omega}(t_n)}{\pi}}u\right)\right]\right\}, \label{mode amplitude near-resonance solution integral in terms of Fresnel integrals}
\end{equation}
where 
\begin{align}
    \mathcal{C}(w)=\int_0^w \cos\left(\frac{\pi w'^2}{2}\right) dw' \label{Fresnel integral C}
    \intertext{and}
    \mathcal{S}(w)=\int_0^w \sin\left(\frac{\pi w'^2}{2}\right) dw' \label{Fresnel integral S}
\end{align}
are the Fresnel integrals. Second, we use the asymptotic behaviour of these integrals to  connect with both \cref{Pre} and \cref{Post}. Third, we confirm that the two solutions we obtain are continuous at the resonance (as they have to be, given that they originate from the near-resonance solution). Fourth, we join the solutions at the resonance by means of the Heaviside step function $H$.

These steps lead to the final, composite expression for the mode solution across resonance:
\begin{equation}
    a_n = {1 \over 2 \omega_n} \frac{Q_n}{\mathcal A_n^2} v_{lm}(t) e^{-im\Phi(t)}  \Bigg\{ { 2\omega_n  \over \omega_n^2 - m^2 \Omega^2}  \mathcal J \left( |z| \right)
      + H(u) \left(\frac{\omega_n}{m\Omega}\right)^{2(l+1)/3}\left( {2 \pi \over m \dot \Omega(t_n)}\right)^{1/2}   e^{- i(\psi - \pi/4)} \Bigg\}.
    \label{composite}
\end{equation}
In this expression
\begin{equation}
    z = \sqrt{{m\dot\Omega(t_n)\over \pi}}u,
\end{equation}
$\mathcal J (z)$ is the (well-behaved) integral 
\begin{equation}
    \mathcal J \left(z \right) = 
    -i\pi  z e^{i\pi z^2/2 } \left\{ \left[ \mathcal C \left(z \right) -{1\over 2} \right]   - i \left[ \mathcal S\left(z\right) - {1\over 2} \right]\right\},
\end{equation}
and the phase $\psi(t)$ is defined as
\begin{equation}
    \psi(t) = \int_{t_n}^t (\omega_n -m\Omega) dt' = \omega_n t - m(\Phi-\Phi_n).
\end{equation}
We need to keep in mind that this expression is valid only for $m\ge 0$, as discussed at the beginning of the calculation. By following exactly the same arguments, it can be shown that the contribution from the $m<0$ resonance (with the same $\omega_n>0$ frequency) is given by $(-1)^m a_n^*$.

The key point is that the expression \eqref{composite} is valid across the resonance and connects to the anticipated pre- and post-resonance behaviour. Moreover, since $\mathcal J(z)$ vanishes at the resonance, the mode amplitude remains non-singular. The regularity of the new resonance solution makes it distinct from the result obtained by \citet{2016PhRvD..94j4028S},  which involves an explicit subtraction of singular terms. This should make our expression more amenable to practical applications. As a first illustration of this, let us work out the corresponding expression for the effective Love number.


\subsection{The Love number}

Given the closed-form expression for the mode amplitude \eqref{composite}, we can easily write down the corresponding expression for the Love number. Following the steps that led to \cref{eq:Love} we see that the contribution from a single mode is given by
\begin{equation}
    k_{lm} 
        = \frac{2 \pi G}{(2 l + 1) R^{2 l + 1}} 
            {1\over v_{lm} } a_{n} Q_n e^{im\Phi},
            \label{klm general}
\end{equation}
which now leads to
\begin{equation}
    k_{lm} = \frac{2 \pi G}{(2 l + 1) R^{2 l + 1}}  {Q_n^2 \over 2\omega_n \mathcal A_n^2}  \Bigg\{ { 2\omega_n  \over \omega_n^2 - m^2 \Omega^2}  \mathcal J \left( |z| \right) 
      + H(u) \left(\frac{\omega_n}{m\Omega}\right)^{2(l+1)/3}  \left( {2 \pi \over m \dot \Omega(t_n)}\right)^{1/2}   e^{- i(\psi - \pi/4)} \Bigg\}
      \label{klmeq}
\end{equation}
(the mode index $n$ is omitted in this section, i.e., $k_{lm}\equiv k_{nlm}$). If we express the result in terms of the static Love number contribution $k_n^0$ from \cref{StaticLove} and include the contribution from the $m<0$ resonance (the complex conjugate), we get
 \begin{equation}
    {k_{l} \over k_n^0} = 
    \Bigg\{ { 2 \omega_n^2  \over \omega_n^2 - m^2 \Omega^2}   \mathcal R \left(|z| \right)  
      + H(u) \left(\frac{\omega_n}{m\Omega}\right)^{2(l+1)/3} \left( { 2 \pi \omega_n^2 \over  m \dot \Omega(t_n)}\right)^{1/2}   \cos\left[ \psi(t)  - {\pi\over 4} \right] \Bigg\},
      \label{Flove}
\end{equation}
where we have introduced the real part
\begin{equation}
    \mathcal R(z) = \mathrm{Re}\ \mathcal J \left(z \right) 
    = \pi z \left\{ \sin\left( {\pi z^2 \over 2} \right) \left[ \mathcal C \left(z \right) -{1\over 2} \right] - \cos \left( {\pi z^2 \over 2} \right)  \left[ \mathcal S\left(z\right) - {1\over 2} \right] \right\} .
\end{equation}

\begin{figure}
    \centering
    \includegraphics[width=\textwidth]{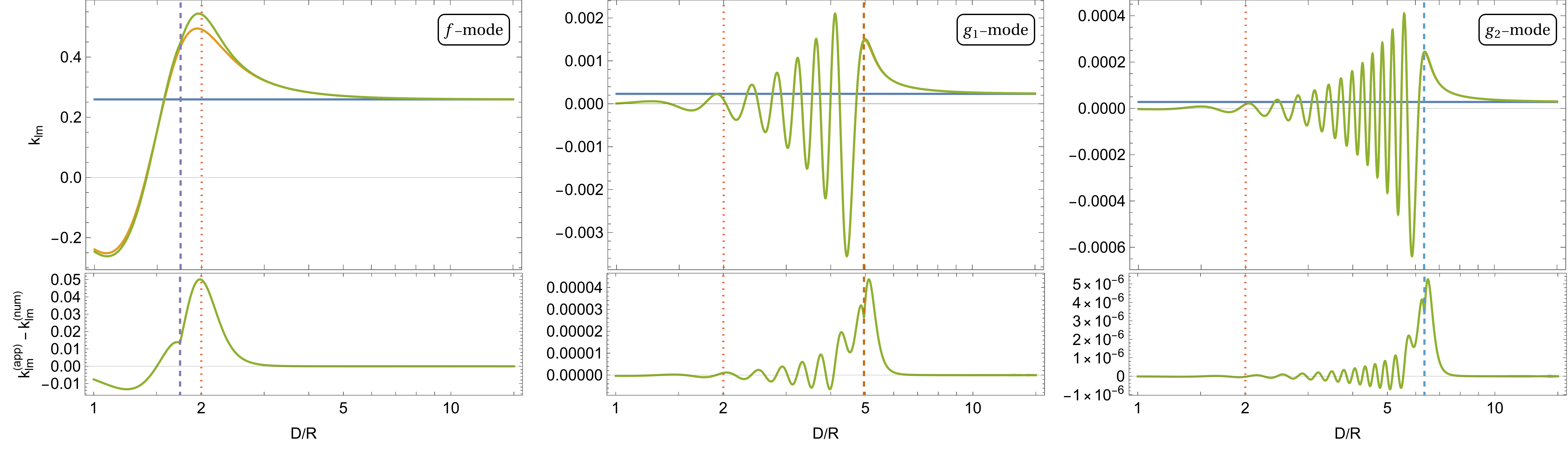}
    \caption{Individual mode contributions to the $l=m=2$ effective Love number, for a polytrope $p\propto\rho^\Gamma$ with $\Gamma=2$ and adiabatic index $\Gamma_1=2.1$. We show results for the real part of the contribution from the fundamental f-mode (left panel) and the first two gravity g-modes (middle and right panels, respectively), based on  using the  mode parameters from Table~\ref{tab:mode_parameters} in \cref{klmeq}. The results illustrate the general behaviour of the solution. The mode contribution to the effective Love number grows as the system approaches resonance. The contribution peaks after resonance and then decays asymptotically while oscillating at the mode frequency. The three examples clearly illustrate that, while the approximate f-mode contribution (in green) differs slightly from the numerical result (orange), our approximate model is very accurate for the (lower frequency) g-modes. This is also reflected in each bottom panel, where we plot the absolute error of the approximate analytical solution, compared to the numerical one. In addition to the mode amplitudes, the contribution of each mode to the static Love number is shown (blue), showing that the contribution to the effective tidal deformability from each mode vanishes asymptotically, as well as vertical lines indicating the location of the mode resonance (dashed line) and the merger distance (here simply taken to be the point when the two stars touch; red dotted line).}
    \label{fig:effective_Love_mode_contributions}
\end{figure}

We compare the result from \cref{Flove} to results obtained from numerical solutions to \cref{mode equation of motion} in Fig.~\ref{fig:effective_Love_mode_contributions}. The accuracy of the approximate solution is evident, with only a slight difference from the numerical result near resonance for the f-mode and excellent agreement through resonance for the first couple of g-modes.

\begin{table}
    \centering
    \begin{tabular}{c|c|c|c}
        \hline\hline
        Mode & $\tilde{\omega}_n$ & $\tilde{Q}_n$ & $k_{n}^0$ \\
        \hline
        f\phantom{$_1$} & 1.2277 & $5.5800\times 10^{-1}$ & $2.5958\times 10^{-1}$ \\
        g$_1$ & 0.2566 & $3.4972\times 10^{-3}$ & $2.3346\times 10^{-4}$ \\
        g$_2$ & 0.1770 & $8.3553\times 10^{-4}$ & $2.8013\times 10^{-5}$ \\
        g$_3$ & 0.1361 & $2.4622\times 10^{-4}$ & $4.1142\times 10^{-6}$ \\
        g$_4$ & 0.1109 & $8.1322\times 10^{-5}$ & $6.7611\times 10^{-7}$ \\
        g$_5$ & 0.0937 & $2.8808\times 10^{-5}$ & $1.1885\times 10^{-7}$ \\
        \hline\hline
    \end{tabular}
    \caption{Eigenfrequencies $\tilde{\omega}_n = \omega_n / \sqrt{GM_\star/R^3}$, mass multipole moments $\tilde{Q}_n = Q_n/M_\star R^2$ and static Love number contributions $k_n^0$ of the f-mode and the first five g-mode overtones with $l=2$, for a polytrope $p\propto\rho^\Gamma$ with $\Gamma=2$ and adiabatic index $\Gamma_1=2.1$.}
    \label{tab:mode_parameters}
\end{table}


\section{The mode energy}

In addition to the evolution of the mode amplitudes, we need to keep track of the  energy deposited in the modes. After all, the rate at which energy is transferred to the modes indicates the level at which a resonance impacts on the orbital evolution \citep{1994MNRAS.270..611L}. In fact, all current discussions of the detectability of low-frequency mode resonances are based on simple energy estimates (see \citealt{2018PhRvD..97b3016A}, \citealt{2022MNRAS.513.4045K}, \citealt{2023PhRvD.108d3003H} and \citealt{2025MNRAS.536.1967C} for recent contributions). If we want to be more precise than this then we need to consider the orbital evolution problem in more detail. As a first step in this direction, we need to work out the evolution of the mode energy.\footnote{%
    We postpone a discussion of the general orbital problem, which involves a considerable amount of additional technical detail, to a companion paper.
}


\subsection{Approximate expression}

The main additional step we need to take in order to determine the mode energy involves working out the time derivative of the amplitude. Keeping in mind that we will (later) need higher derivatives to work out the rate  of gravitational-wave emission, it makes sense to design an efficient way of dealing with this problem. Luckily, this is straightforward. Starting from the equation for the mode amplitude \eqref{mode equation of motion} and introducing  $c_n = \dot a_n$ we have 
\begin{equation}
    \ddot{c}_n + \omega_n^2 c_n = {GM' W_{lm} Q_n \over \mathcal A_n^2} {d\over dt}\left[  \frac{1}{D^{l+1}} e^{-im\Phi} \right] \approx {GM' W_{lm} Q_n \over \mathcal A_n^2} \frac{(-im\Omega)}{D^{l+1}} e^{-im\Phi}  = {V_{lm} Q_n \over \mathcal A_n^2} e^{-im\Phi},
\end{equation}
where the last equality defines the function $V_{lm}(t)$. Comparing to the equation we solved for the amplitude, i.e. \cref{mode equation of motion}, and recalling the steps in the derivation of the resonance solution \eqref{composite}, we can  write down a closed-form approximation also for $c_n$. This argument can be extended to higher derivatives as required.  

The  problems for $a_n$ and $c_n$ are clearly identical once we replace $v_{lm}$ with $V_{lm}$. Going back to \cref{composite} we infer that 
\begin{equation}
    c_n = {1 \over 2 \omega_n} \frac{Q_n}{\mathcal A_n^2} V_{lm}(t)  e^{-im\Phi(t)} \Bigg\{ { 2\omega_n  \over \omega_n^2 - m^2 \Omega^2}  \mathcal J \left( |z| \right)
      + H(u) {V_{lm}(t_n) \over V_{lm}(t)} \left( {2 \pi \over m \dot \Omega(t_n)}\right)^{1/2}   e^{- i(\psi - \pi/4)} \Bigg\}.
      \label{csolution}
\end{equation}
With 
\begin{equation}
    V_{lm} = - im \Omega v_{lm} \Longrightarrow V_{lm}(t_n) = - i \omega_n v_{lm}(t_n),
\end{equation}
we then get
\begin{equation}
    \dot a_n = c_n \approx {1 \over 2 \omega_n} \frac{Q_n}{\mathcal A_n^2}  e^{-im\Phi(t)}  \Bigg\{ { 2\omega_n  \over \omega_n^2 - m^2 \Omega^2}  [-im\Omega v_{lm}(t)]  \mathcal J \left(|z| \right)
      + H(u) [-i\omega_n v_{lm}(t_n)]\left( {2 \pi \over m \dot \Omega(t_n)}\right)^{1/2}   e^{- i(\psi - \pi/4)} \Bigg\},
\end{equation}
which allows us to work out the mode energy.

After some algebra, we arrive at 
\begin{multline}
    E_n =  
      {1 \over 2} \frac{Q_n^2}{\mathcal A_n^2}  {v_{lm}^2(t) \over \omega_n^2} 
      \Bigg\{  {\omega_n^2(\omega_n^2 + m^2\Omega^2)  \over (\omega_n^2 - m^2 \Omega^2)^2}  \mathcal J ^2 +  H(u)  {\pi \omega_n^2  \over m \dot \Omega(t_n)} \left[ { v^2_{lm}(t_n) \over v_{lm}^2(t)}\right]
    \\ +  H(u) 
       {\omega_n(\omega_n+m\Omega )\over \omega_n^2 - m^2 \Omega^2}  \left[   {v_{lm}(t_n) \over v_{lm}(t)}\right]   \left[ {2 \pi \omega_n^2 \over m^2\dot \Omega(t_n) }\right]^{1/2} \left[ \mathcal R   \cos \left(   \int_{t_n}^t (\omega_n -m\Omega) dt  - \pi/4\right)  - \mathcal I   \sin \left(  \int_{t_n}^t (\omega_n -m\Omega) dt  - \pi/4\right) \right] \Bigg\}  ,
       \label{compEn}
\end{multline} 
where we have introduced
\begin{equation}
    \mathcal I(z) = \mathrm{Im}\ \mathcal J(z) 
    = - \pi z \left\{   \sin\left( {\pi z^2 \over 2} \right) \left[ \mathcal S \left(z \right) -{1\over 2} \right] + \cos \left( {\pi z^2 \over 2} \right)  \left[ \mathcal C\left(z\right) - {1\over 2} \right]  \right\}
\end{equation}
and
\begin{equation}
    \mathcal J^2 = \pi^2 z^2 \left[ \left( \mathcal C(z)-{1\over 2} \right)^2 + \left( \mathcal S(z) - {1\over 2} \right)^2\right].
\end{equation}
In \cref{compEn} we have chosen to write the energy in such a way that the pre-resonance (equilibrium tide) energy is factored out.

\begin{figure}
    \centering
    \includegraphics[width=\textwidth]{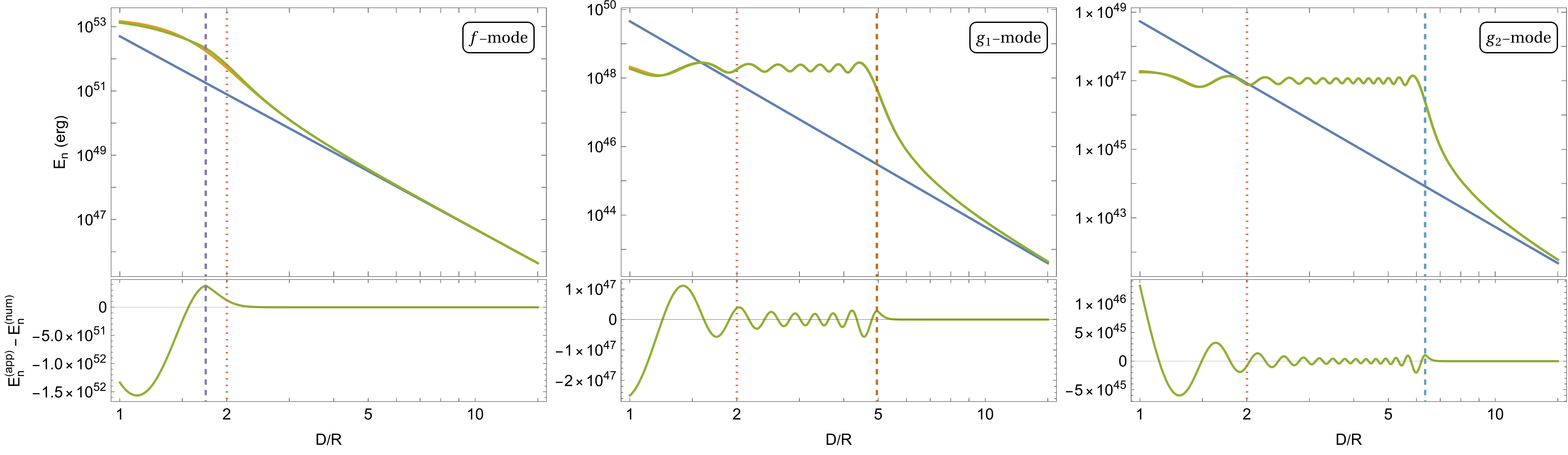}
    \caption{The mode energy obtained from \cref{compEn} for the f-mode (left panel) and the first two g-modes (middle and right panels) for the same stellar model as in Fig.~\ref{fig:effective_Love_mode_contributions} and $l=m=2$. The corresponding mode contribution to the energy associated with the static tide (obtained from \cref{pre_energy} with $\Omega=0$) is also included (in blue), for comparison. Comparing the numerical results (orange) to the approximate solution (green) we see that the error in the energy is similar to that of the amplitudes. As in Fig.~\ref{fig:effective_Love_mode_contributions}, the absolute error of the approximate analytical solution, compared to the numerical one, is depicted in the bottom panel, whereas the vertical lines indicate the location of the mode resonance (dashed) and the merger distance (red dotted). The results help motivate the common assumption that the behaviour can be modelled as a more or less instantaneous transfer of energy at resonance.}
    \label{fig:energy_mode_contributions}
\end{figure}

Again, a comparison with numerical results shows that this approximate expression is robust. Given the accuracy of the mode amplitudes, the results provided for the energy  in Fig.~\ref{fig:energy_mode_contributions} are, perhaps, not surprising. However, they do provide confidence in the strategy we adopted for the time derivative of the amplitude. This will be important if we go further and work out, say, the rate of gravitational-wave emission which requires the third time derivative.

\begin{table}
    \centering
    \begin{tabular}{c|c|c|c}
        \hline\hline
        Mode & $\tilde{\omega}_n$ & $\tilde{Q}_n$ & $k_{n}^0$ \\
        \hline
        f & 1.3896 & $4.5400\times 10^{-1}$ & $1.3414\times 10^{-1}$ \\
        i & 0.4538 & $2.1700\times 10^{-2}$ & $2.8739\times 10^{-3}$ \\
        \hline\hline
    \end{tabular}
    \caption{Eigenfrequencies $\tilde{\omega}_n = \omega_n / \sqrt{GM_\star/R^3}$, mass multipole moments $\tilde{Q}_n = Q_n/M_\star R^2$ and static Love number contributions $k_n^0$ of the f-mode and the i-mode with $l=2$, for one of the models considered in \citet{interface_paper}.}
    \label{tab:mode_parameters_interface_paper_model}
\end{table}

\begin{figure}
    \newlength{\energyfigureheight}
    \setlength{\energyfigureheight}{5.6cm}
    \setlength{\tabcolsep}{0pt}
    \centering
    \begin{tabular}{cc}
        \includegraphics[height=\energyfigureheight]{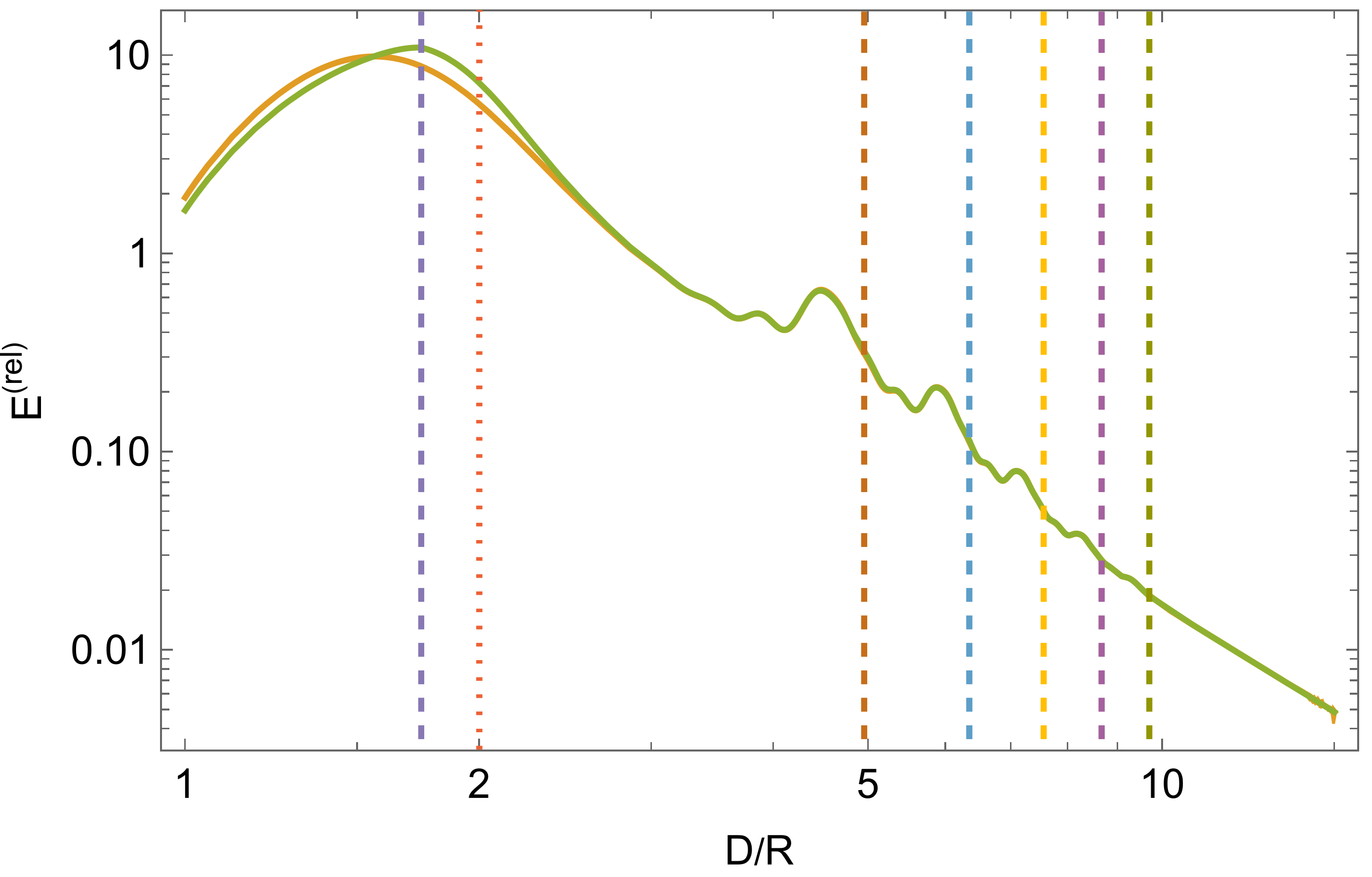} & \includegraphics[height=\energyfigureheight]{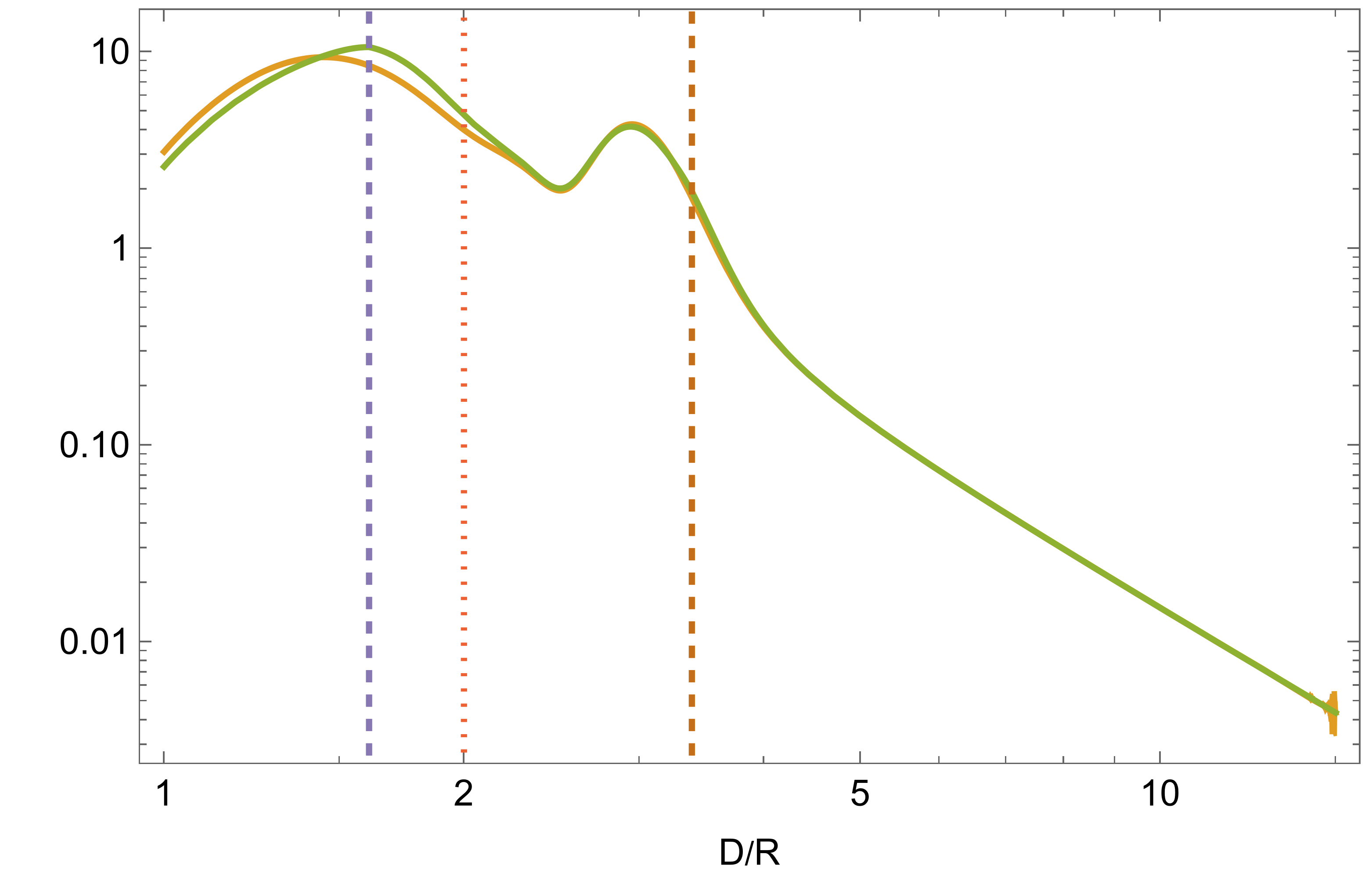}
    \end{tabular}
    \caption{Summed mode contributions to the energy obtained from \cref{compEn}, evaluated relative to the static tide energy (i.e. subtracting the static tide energy obtained from \cref{pre_energy} with $\Omega=0$ and then dividing by it). The approximate (green) and numerical (orange) solutions are shown, for two different stellar models and $l=m=2$. The left panel corresponds to the polytropic model from Figs.~\ref{fig:effective_Love_mode_contributions} and \ref{fig:energy_mode_contributions}, for all the modes listed in Table~\ref{tab:mode_parameters}. As in the previous plots, the vertical dashed lines indicate the location of the various mode resonances, starting from the g$_5$-mode resonance on the right and ending up with the f-mode resonance on the left, right after merger (red dotted vertical line). The results indicate the modest impact of the low-order g-modes, which may nevertheless be marginally detectable with next-generation gravitational-wave instruments \citep[see, for example,][]{2025MNRAS.536.1967C}. The right panel corresponds to one of the models from \citet{interface_paper} where an interface mode associated with a first-order phase transition contributes prominently. The interface mode resonance (brown dashed line) is followed by the merger (red dotted line), with the f-mode resonance (purple dashed line) again occurring just after merger. The mode parameters of this model are given in Table~\ref{tab:mode_parameters_interface_paper_model}.}
    \label{fig:relative_energy}
\end{figure}

\begin{figure}
    \newlength{\Lovefigureheight}
    \setlength{\Lovefigureheight}{5.55cm}
    \setlength{\tabcolsep}{0pt}
    \centering
    \begin{tabular}{cc}
        \includegraphics[height=\Lovefigureheight]{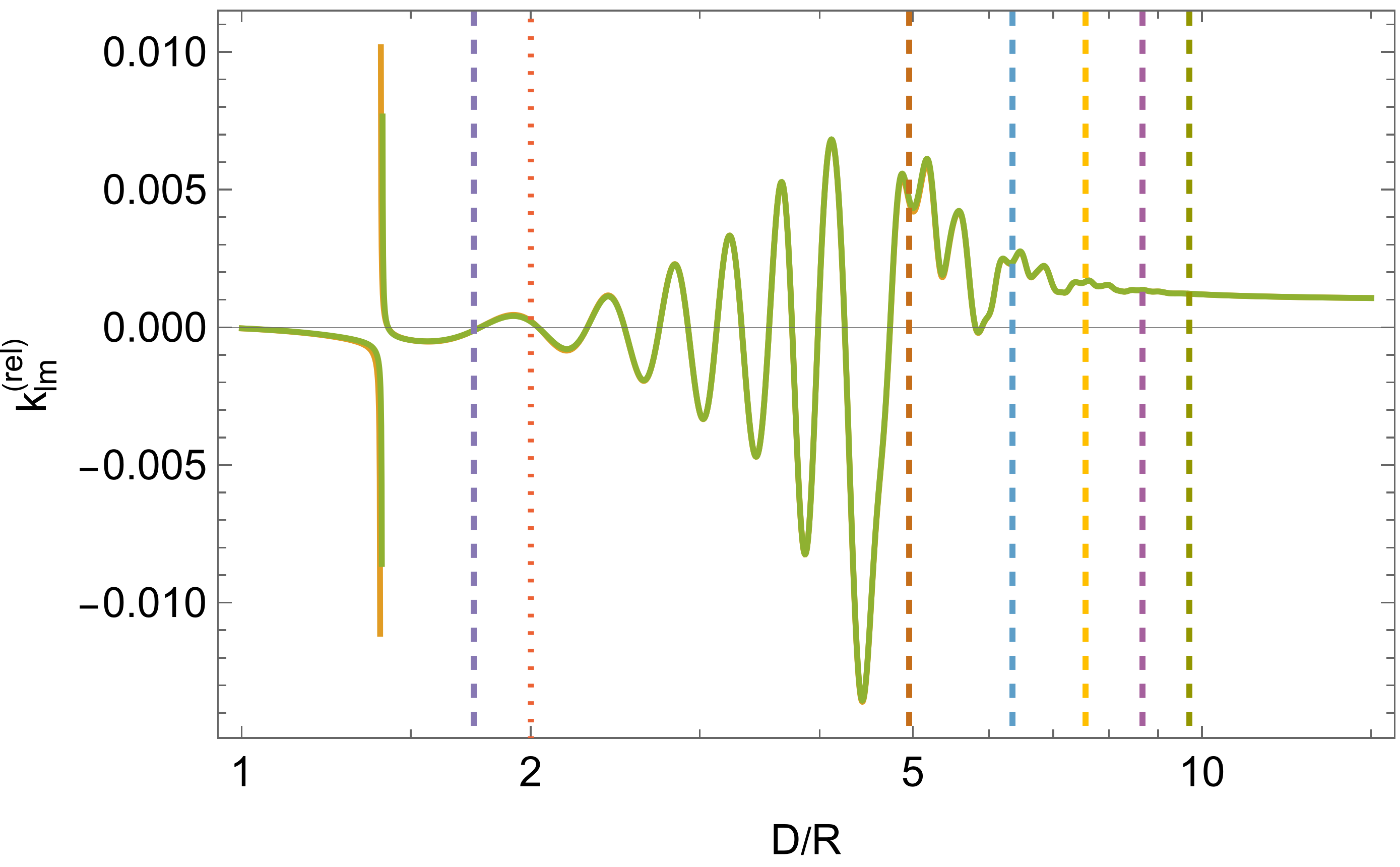} & \includegraphics[height=\Lovefigureheight]{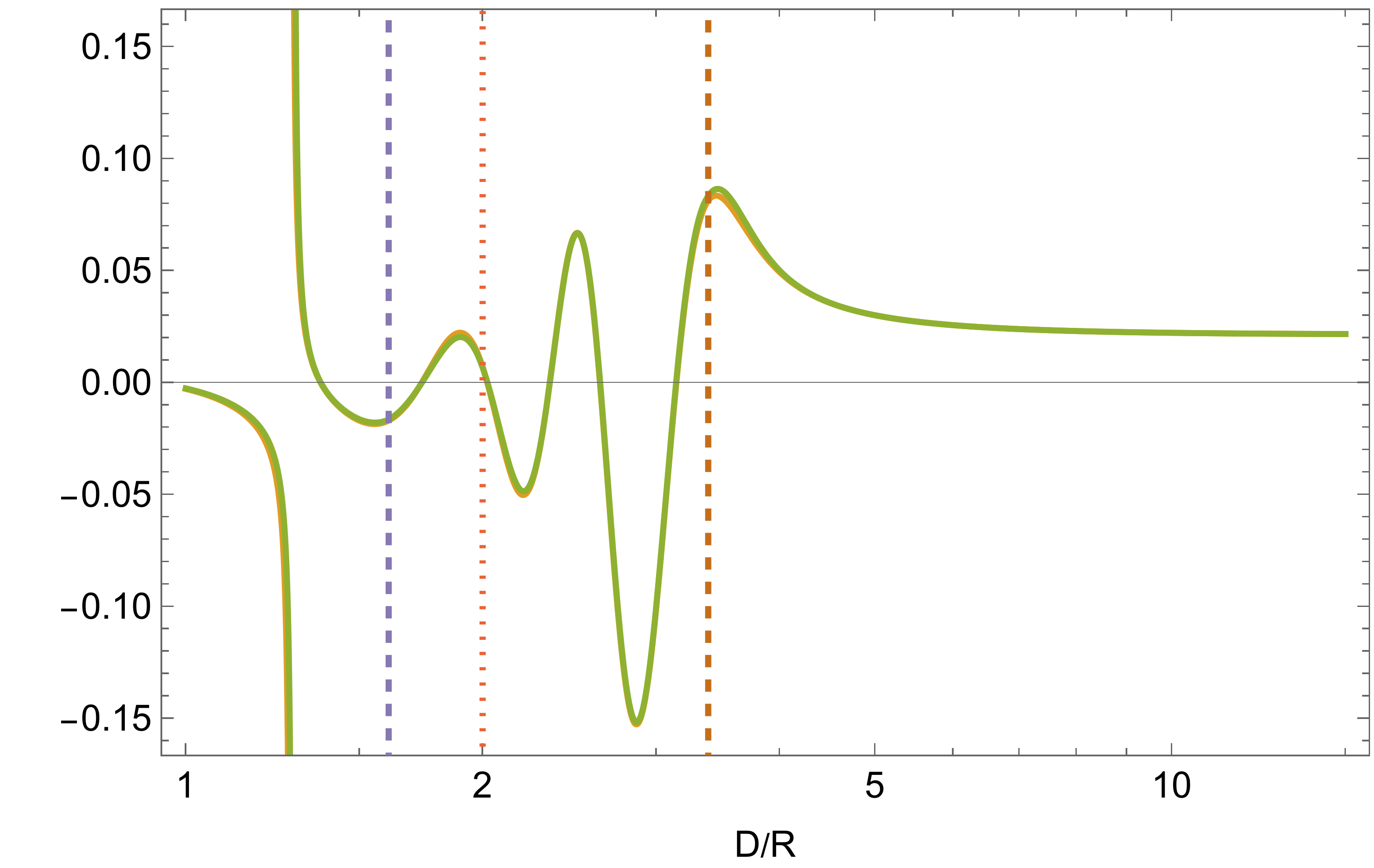}
    \end{tabular}
    \caption{Summed mode contributions to the (real part of the)  effective Love number from \cref{klmeq}, evaluated relative to the f-mode contribution (i.e. subtracting the f-mode contribution and then dividing by it). The approximate (green) and numerical (orange) solutions are shown, for the same models as in Fig.~\ref{fig:relative_energy} and $l=m=2$. The modest impact of the low-order g-modes can be seen here too, compared to the larger by an order of magnitude contribution of the interface mode. The divergence of the solution after the f-mode resonance is due to the f-mode Love number contribution crossing zero (see Fig.~\ref{fig:effective_Love_mode_contributions}).}
    \label{fig:relative_Love_number}
\end{figure}


\subsection{Simple energy estimate}

The impact on the gravitational-wave phasing associated with individual mode resonances is often assessed (see, for instance, \citealt{2018PhRvD..97b3016A}, \citealt{2022MNRAS.513.4045K}, \citealt{2023PhRvD.108d3003H} and \citealt{2025MNRAS.536.1967C}) using a simple energy estimate. Following \cite{1994MNRAS.270..611L}, the energy involved is taken to be given by the second term in \cref{compEn}, which is constant after the resonance. In essence, this model assumes that this energy is instantaneously transferred from the orbit at resonance. This assumption is motivated by the behaviour illustrated in \cref{fig:energy_mode_contributions}, but obviously leaves out the contribution from the equilibrium tide (before and after resonance) and the energy in the free oscillations of the star.

Assuming that the additional energy loss is associated with a distinct frequency, the change in the number of gravitational-wave cycles is given by 
\begin{equation}
    \Delta \mathcal N \approx - \int_{f_a}^{f_b} f {d E_\mathrm{tide} \over dD} {1 \over \dot E_\mathrm{orb} }  dD.
\end{equation}
After integration, this leads to 
\begin{equation}
    \Delta \mathcal  N \approx -  \left( {f \Delta E_\mathrm{tide} \over \dot E_\mathrm{orb} } \right)_{f=f_n} \approx - \left( {3 \over 2} {f t_D \Delta E_\mathrm{tide} \over E_\mathrm{N} } \right)_{f=f_n},
    \label{dNres}
\end{equation}
where $\Delta E_\mathrm{tide}$ is the total energy transferred from the orbit to the resonant mode, $E_\mathrm{N}$ is the (Newtonian) orbital energy, $t_D$ is the orbital evolution timescale \eqref{timeD} and the expression should be evaluated at the resonance frequency $f=f_n$. This then allows us to estimate the induced change in the gravitational-wave phase and assess to what extent the resonance we have considered may be interesting from an observational perspective. 

Specifically, from \cref{compEn} we have
\begin{equation}
    E_n =  \frac{Q_{n}^2 }{2\mathcal A_n^2} \left[ \frac{GM'W_{lm}}{D^{l+1}(t_n)}\right]^2 \left( { \pi \over m \dot \Omega(t_n)}\right),
\end{equation}
where the orbital parameters are evaluated at resonance. With $W_{2\pm 2} = \sqrt{3\pi/10}$, assuming the leading-order (quadrupole) inspiral rate and taking the orbit to be Keplerian, we have
\begin{equation}
    \Omega^2 = {GM\over D^3} \longrightarrow D \approx \left( {m\over \tilde \omega_n}\right)^{2/3} \left( {M\over M_\star} \right)^{1/3} R,
\end{equation}
and the energy associated with the two $|m|=2$ modes (which contribute equally because of the symmetries of the problem) becomes
\begin{equation}
    E_n  =   {\pi^2 \over 512}  \tilde \omega_n^{1/3}  \tilde Q_{n}^2 
      \left( {Rc^2 \over GM_\star}\right)^{3/2} q \left( { 2\over   1+q } \right)^{5/3} M_\star c^2,
\end{equation}
where $q = M'/M_\star$ is the mass ratio. This result accords with the expression used by, for example,  \cite{2023PhRvD.108d3003H} and \cite{2025MNRAS.536.1967C}.

This energy is assumed to be extracted from the orbit at the resonance frequency (for $m=2$)
\begin{equation}
    \omega_n = 2\pi f_n = 2\Omega =  2\pi f.
\end{equation}
This corresponds to an orbital separation 
\begin{equation}
    D_n = \left[ {4GM_\star(1+q)\over \omega_n^2} \right]^{1/3},
\end{equation}
and it readily follows that the (Newtonian) orbital energy is
\begin{equation}
    E_\mathrm{N} =  - {1\over 2^{5/3}} \left( {GM_\star \over R c^2} \right) \tilde \omega_n^{2/3} { q \over (1+q)^{1/3}} M_\star c^2.
\end{equation}
Finally, \cref{dNres} leads to
\begin{equation}
    \Delta \mathcal N \approx  - 4\times10^{-4} \tilde f_n^{-2} \tilde Q_n^2  \left( {c^2 R \over GM_\star}\right)^5 {1\over q (1+q)}.
    \label{dNnew}
\end{equation}
which is the result used to assess the mode detectability in \citet{2025MNRAS.536.1967C,interface_paper}.


\section{Final remarks} \label{sec:final remarks}

In conclusion, we have provided an approximate, closed-form, solution for the amplitude of a resonant stellar oscillation mode in a binary system. Compared to numerical solutions for the mode amplitude, the new model performs well in all relevant domains. In particular, it remains valid as the system passes through resonance. In terms of accuracy, the model is comparable to the previous matched asymptotics expressions from \citet{2016PhRvL.116r1101H} and \citet{2016PhRvD..94j4028S}. In fact, the current model is more amenable to applications as it does not involve an explicit subtraction of terms that become singular at resonance. Similar results have been obtained by \citet{2024PhRvD.110b4039Y}, who derived analytical approximations for the equilibrium and dynamical pieces of the driven mode response, following a somewhat different logic. A direct comparison between the two models, involving a reformulation of the expressions derived here in phase space (see \cref{sec:phase space}), shows that the formulae of \citet{2024PhRvD.110b4039Y} perform as well as the ones presented here for g-modes, but are less reliable for the f-mode. The difference likely arises, as pointed out by \citet{2024PhRvD.110b4039Y}, because the f-mode induces a significant back-reaction on the orbit and the impact of this is treated differently in the two calculations. These comparisons establish confidence in the results.

Given the expressions we have provided, the new approximate solution for the resonant mode amplitudes is readily---once we combine it with results for mode frequencies and mass multipole moments---used to provide a mode-sum representation for the effective tidal deformability. As as example of this, we show the summed energies for two different stellar models in \cref{fig:relative_energy}. The summed effective Love number is shown, for the same two models, in \cref{fig:relative_Love_number}.

Evidently, the results we have provided allow us to make progress on assessing the relevance of low-frequency mode resonances that may leave an observable imprint on gravitational-wave signals from binary inspirals involving neutron stars. However, in terms of the impact of tides on the evolution of a binary system, our results only address the response of the star to the presence of the companion. In order to build a complete model, we need to consider also the ``outer problem''; how the tidal excitation impacts on the binary orbit and its evolution. From a conceptual point of view, the steps one would have to take to complete such a model are clear. Taking the solution for the static tide problem from \citet{2008PhRvD..77b1502F} as a guide, one would have to i) work out how the excited modes impact on the equilibrium orbit of the system (used in the adiabatic evolution), ii) quantify how this changes the energy of the system, and iii) estimate the rate of gravitational-wave emission (via the standard post-Newtonian multipole formulas). Once this is done, the mode-sum representation for the tide, including resonances, would be developed to the same level as current models for the static tidal deformability. \citet{2024PhRvD.110b4039Y} have already taken important steps in this direction by studying the coupled dynamics between the orbit and the tide, focussing on the contribution of the f-mode \citep[see also][]{2025PhRvD.111h4029Y}. Our specific aim here was to provide an accurate framework for the description of each mode response, including after resonance. This regime is particularly relevant for low-frequency g-modes and hence our results represent a notable step forward for the modelling required to establish to what extent next-generation gravitational-wave instruments will be able to probe neutron star physics beyond mass and radius. The results we have discussed---especially for the mode energy---prepare the ground for the required models, but there is still a fair amount of work to be done. We expect to report on relevant progress in the near future.


\section*{Acknowledgements}

The authors would like to thank Hang Yu for helpful discussions and useful suggestions.
PP acknowledges support from the Mar\'ia Zambrano Fellowship Programme (ZAMBRANO21-22), funded by the Spanish Ministry of Universities and the University of Alicante through the European Union's ``Next Generation EU'' package, as well as from the grant PID2021-127495NB-I00, funded by MCIN/AEI/10.13039/501100011033 and by the European Union, from the Astrophysics and High Energy Physics programme of the Generalitat Valenciana ASFAE/2022/026, funded by the Spanish Ministry of Science and Innovation (MCIN) and the European Union's ``Next Generation EU'' package (PRTR-C17.I1), and from the Prometeo 2023 excellence programme grant CIPROM/2022/13, funded by the Ministry of Innovation, Universities, Science, and Digital Society of the Generalitat Valenciana. 
NA gratefully acknowledges support from the STFC via grant No.~ST/Y00082X/1.
FG acknowledges funding from the European Union’s Horizon Europe research and innovation programme under the Marie Sk{\l}odowska-Curie grant agreement No.~101151301.

Views and opinions expressed are however those of the authors only and do not necessarily reflect those of the European Union or the European Research Council. Neither the European Union nor the granting authority can be held responsible for them.


\section*{Data Availability}

Additional data related to this article will be shared on reasonable request to the corresponding author.


\bibliographystyle{mnras}
\bibliography{bibliography}


\appendix


\section{The derivation of the composite mode solution} \label{sec:derivation}

The derivation of the closed-form mode solution \eqref{composite} involves the following steps. First, we rewrite the integral from \cref{nearsol} in terms of the Fresnel integrals $\mathcal{C}$ and $\mathcal{S}$ as
\begin{equation}
	\int_{-\infty}^u e^{-i(m/2)\dot{\Omega}(t_n){u'}^2} du' = \sqrt{\frac{\pi}{m\dot{\Omega}(t_n)}}\left\{\frac{1}{2}+\mathcal{C}\left(\sqrt{\frac{m\dot{\Omega}(t_n)}{\pi}}u\right)-i\left[\frac{1}{2}+\mathcal{S}\left(\sqrt{\frac{m\dot{\Omega}(t_n)}{\pi}}u\right)\right]\right\},
\end{equation}
where
\begin{align}
    \mathcal{C}(w)=\int_0^w \cos\left(\frac{\pi w'^2}{2}\right) dw'
    \intertext{and}
    \mathcal{S}(w)=\int_0^w \sin\left(\frac{\pi w'^2}{2}\right) dw' .
\end{align}
It is important to note that  $\mathcal C$ and $\mathcal S$  are both odd functions. Then we use the asymptotic behaviour
\begin{equation}
    \mathcal C(\lambda) \approx {1\over 2} + {1\over \pi \lambda } \sin\left(\frac{\pi \lambda^2}{2}\right)
\end{equation}
and
\begin{equation}
    \mathcal S(\lambda) \approx {1\over 2} - {1\over \pi \lambda } \cos\left(\frac{\pi \lambda^2}{2}\right) , 
\end{equation}
to  make contact with both \cref{Pre} and \cref{Post}, adapting the logic outlined for the Gauss error function in the main text.

For the post-resonance part of the solution, we start from
\begin{equation}
	a_n =
     {i \over 2 \omega_n} \frac{Q_n}{\mathcal A_n^2} v_{lm}(t_n) e^{i(m/2)\dot{\Omega}(t_n)u^2} I \left(\sqrt{{m\dot\Omega(t_n)\over \pi}}u \right) e^{-im\Phi(t)} \ ,
\end{equation}
where
\begin{equation}
     I(\lambda) = \left( {\pi \over m \dot \Omega(t_n)}\right)^{1/2}  \left\{ \sqrt{2} e^{-i\pi/4}  + \left[ \mathcal C(\lambda) -{1\over 2} \right]   - i \left[ \mathcal S(\lambda) - {1\over 2} \right] \right\} \ .
\end{equation}
This leads to
\begin{multline}
    a_n =
     {i \over 2 \omega_n} \frac{Q_n}{\mathcal A_n^2} v_{lm}(t_n) e^{i(m/2)\dot{\Omega}(t_n)u^2-im\Phi(t)} \left( {\pi \over m \dot \Omega(t_n)}\right)^{1/2}   \sqrt{2} e^{-i\pi/4} \\
     + {i \over 2 \omega_n} \frac{Q_n}{\mathcal A_n^2} v_{lm}(t_n) e^{i(m/2)\dot{\Omega}(t_n)u^2} \left( {\pi \over m \dot \Omega(t_n)}\right)^{1/2} 
     \left\{ \left[ \mathcal C\left(\sqrt{{m\dot\Omega(t_n)\over \pi}}u\right) -{1\over 2} \right] - i \left[ \mathcal S\left(\sqrt{{m\dot\Omega(t_n)\over \pi}}u\right) - {1\over 2} \right]\right\} e^{-im\Phi(t)} \ .
\end{multline}
In the first term, we undo the near-resonance Taylor expansion (as in \cref{subsec:near_resonance_solution,subsec:beyond_resonance}) and in the second term we simply change $v_{lm}(t_n)\to v_{lm}(t)$ in order to match the equilibrium tide solution away from resonance. We do not make the same change in both terms as we know that the free mode oscillations should be at constant amplitude after the resonance. This leads to
\begin{multline}
    a_n =
     {1 \over 2 \omega_n} \frac{Q_n}{\mathcal A_n^2} v_{lm}(t_n)  \left( {2 \pi \over m \dot \Omega(t_n)}\right)^{1/2}   e^{- i(\omega_n t+ m\Phi_n - \pi/4)} \\
     + {i \over 2 \omega_n} \frac{Q_n}{\mathcal A_n^2} v_{lm}(t) \left( {\pi \over m \dot \Omega(t_n)}\right)^{1/2} 
     e^{i(m/2)\dot{\Omega}(t_n)u^2}  \left\{ \left[ \mathcal C\left(\sqrt{{m\dot\Omega(t_n)\over \pi}}u\right) -{1\over 2} \right] - i \left[ \mathcal S\left(\sqrt{{m\dot\Omega(t_n)\over \pi}}u\right) - {1\over 2} \right]\right\} e^{-im\Phi(t)} \ .
\end{multline}
Further guided by the asymptotics, we rewrite this as
\begin{multline}
    a_n =
     {1 \over 2 \omega_n} \frac{Q_n}{\mathcal A_n^2} v_{lm}(t_n)  \left( {2 \pi \over m \dot \Omega(t_n)}\right)^{1/2}   e^{- i(\omega_n t+ m\Phi_n - \pi/4)} \\
     + {i \over 2 \omega_n} \frac{Q_n}{\mathcal A_n^2} v_{lm}(t) \left( {\pi \over m \dot \Omega(t_n)}\right)^{1/2} \left( {m\dot \Omega(t_n) u \over m\dot \Omega(t_n) u} \right) 
     e^{i(m/2)\dot{\Omega}(t_n)u^2}  \left\{ \left[ \mathcal C\left(\sqrt{{m\dot\Omega(t_n)\over \pi}}u\right) -{1\over 2} \right] - i \left[ \mathcal S\left(\sqrt{{m\dot\Omega(t_n)\over \pi}}u\right) - {1\over 2} \right]\right\} e^{-im\Phi(t)} \ ,
\end{multline}
and then undo the Taylor expansion in the denominator using \cref{Omndot} to get
\begin{equation}
    a_n =
     {1 \over 2 \omega_n} \frac{Q_n}{\mathcal A_n^2} v_{lm}(t_n)  \left( {2 \pi \over m \dot \Omega(t_n)}\right)^{1/2}   e^{- i(\omega_n t+ m\Phi_n - \pi/4)} 
     + \frac{Q_n}{\mathcal A_n^2} v_{lm}(t)  { 1  \over \omega_n^2 - m^2 \Omega^2}  \mathcal J \left( \sqrt{{m\dot\Omega(t_n)\over \pi}}u \right)
      e^{-im\Phi(t)} \ .
\end{equation}
With
\begin{equation}
    z = \sqrt{{m\dot\Omega(t_n)\over \pi}}u \ , 
\end{equation}
the integral we need is
\begin{equation}
    \mathcal J \left(z \right) = 
    -i\pi  z e^{i\pi z^2/2 } 
    \left\{ \left[ \mathcal C \left(z \right) -{1\over 2} \right]   - i \left[ \mathcal S\left(z\right) - {1\over 2} \right]\right\} \ ,
\end{equation}
which is well-behaved. 

For the pre-resonance behaviour, we need to massage the solution differently. In this case, when $u<0$, the asymptotic behaviour suggests that we start from
\begin{multline}
    a_n = - {i \over 2 \omega_n} \frac{Q_n}{\mathcal A_n^2} v_{lm}(t) \left( {\pi \over m \dot \Omega(t_n)}\right)^{1/2} \left({m\dot \Omega(t_n) u \over m\dot \Omega(t_n) u} \right) \\
     \times
     e^{i(m/2)\dot{\Omega}(t_n)u^2}  \left\{ \left[ \mathcal C\left(\sqrt{{m\dot\Omega(t_n)\over \pi}}|u|\right) -{1\over 2} \right]   - i \left[ \mathcal S\left(\sqrt{{m\dot\Omega(t_n)\over \pi}}|u|\right) - {1\over 2} \right]\right\} e^{-im\Phi(t)} \ .
\end{multline}
In this case (again, undoing the Taylor expansion in the denominator, like before), the amplitude becomes
\begin{equation}
    a_n =
      \frac{Q_n}{\mathcal A_n^2} v_{lm}(t)  { 1  \over \omega_n^2 - m^2 \Omega^2}  \mathcal J \left( \sqrt{{m\dot\Omega(t_n)\over \pi}} |u| \right)
      e^{-im\Phi(t)} \ .
\end{equation}

Finally, making use of
\begin{equation}
	v_{lm}(t_n) = \frac{GM'W_{lm}}{D^{l+1}(t_n)} \approx GM'W_{lm} \left( {\Omega^2 (t_n) \over GM} \right)^{(l+1)/3} 
    \approx v_{lm}(t)\left(\frac{\omega_n}{m\Omega(t)}\right)^{2(l+1)/3} \ , 
\end{equation}
and the Heaviside function, $H(u)$, we arrive at the composite solution
\begin{equation}
    a_n = {1 \over 2 \omega_n} \frac{Q_n}{\mathcal A_n^2} v_{lm}(t)  \Bigg\{ { 2\omega_n  \over \omega_n^2 - m^2 \Omega^2}  \mathcal J \left( \sqrt{{m\dot\Omega(t_n)\over \pi}}|u| \right)
      e^{-im\Phi(t)} 
      + H(u) \left(\frac{\omega_n}{m\Omega}\right)^{2(l+1)/3}\left( {2 \pi \over m \dot \Omega(t_n)}\right)^{1/2}   e^{- i(\omega_n t+ m\Phi_n - \pi/4)} \Bigg\} \ ,
    \label{composite 2}
\end{equation}
which is \cref{composite} in the main text. The first term captures the driving of the mode by the tide throughout the inspiral, whereas the second one corresponds to the post-resonance free oscillation of the mode. $\mathcal{J}(|z|)$ is of order unity before and after the resonance, and goes smoothly to zero at the resonance. Although the limit value of the first term at resonance is finite, it changes sign due to $(\omega_n^2 - m^2 \Omega^2)$ passing through zero. This should not be a problem, however; when the first term reaches its resonance value, the post-resonance mode oscillation is turned on, thereby guaranteeing the continuity of the solution at the resonance.


\section{Adding rotation: The phase space formalism} \label{sec:phase space}

In the presence of rotation, the usual mode decomposition \eqref{mode sum} generally leads to the mode equations of motion coupling to one another, which practically makes the problem untractable. To circumvent this, one should instead employ a phase space expansion (see \citealt{2001PhRvD..65b4001S} for the relevant details), in which the displacement and its derivative are decomposed simultaneously in terms of the basis vectors $(\xi^i_n,\,i\omega_n\xi^i_n)$ as follows:
\begin{equation}
    \begin{bmatrix}
        \xi^i(t, x^i) \\
        \partial_t \xi^i(t, x^i)
    \end{bmatrix}
        = \sum_n \left\{ 
        c_n(t) 
    \begin{bmatrix}
        \xi_n^i(x^i) \\
        i \omega_n \xi_n^i(x^i)
    \end{bmatrix}
        + c_n^*(t) 
    \begin{bmatrix}
        \xi_n^{i *}(x^i) \\
        - i \omega_n \xi_n^{i *}(x^i)
    \end{bmatrix}
        \right\}.
        \label{phase space expansion}
\end{equation}
Using the phase-space amplitudes $c_n$ and assuming that the orbital angular momentum is aligned with the stellar spin, we obtain the first-order equation of motion in the rotating frame as
\begin{equation}
	\dot{c}_n-i\omega_n c_n=-\frac{i}{\mathcal{B}_n} \sum_l v_{lm}(t) Q_{nl} e^{-im[\Phi(t)-\Omega_s t]} \ ,
	\label{phase amplitude equation of motion}
\end{equation}
where $\Omega_s$ is the angular velocity of the star, the constant $\mathcal{B}_n$ is defined through the mode orthogonality condition, generalised for rotating stars, and the sourcing term no longer corresponds to a single multipole degree $l$.
The Love number may then be expressed as
\begin{equation}
    k_{lm}=\frac{2\pi G}{(2l+1)R^{2l+1}}\frac{e^{im[\Phi(t)-\Omega_s t]}}{v_{lm}}
        \sum_{n'}\left[Q_{n'_+ l} \, c_{n'_+}+(-1)^m Q_{n'_- l} \, c_{n'_-}^*\right],
    \label{Love number in phase space with rotation}
\end{equation}
with the label $n'$ representing the modes with a given azimuthal order $m$ and the $(+,-)$ subscripts dictating the sign of $m$ inside the corresponding quantity (e.g., $Q_{n'_- l}$ is the mass multipole moment of the $l$-component of a mode with an azimuthal number equal to $-m$; for the detailed steps and definitions, see \citealt{2024MNRAS.527.8409P}). Furthermore, the expression for the mode energy is
\begin{equation}
	E_n=|c_n|^2\omega_n\mathcal{B}_n.
	\label{mode energy in phase space with rotation}
\end{equation}

Even though the phase-space formalism results in elegant and exact analytical expressions, their use in actual applications is far from trivial, due to complications introduced into the problem by rotation. To begin with, the background star is no longer spherical, which introduces a polar dependence into the equilibrium quantities. Second, the induced coupling between different multipoles complicates the angular dependence of a mode, which can no longer be described by a single spherical harmonic. Third, it splits the modes corresponding to the same multipole degree $l$ but to different azimuthal orders $m$, which are degenerate in the non-rotating case. Finally, a new mode family emerges, the inertial modes, which mix with the normal fluid modes, introducing ambiguities in the description of a mode. For these reasons, the price that is usually paid is the use of the slow-rotation approximation, in which rotation is treated perturbatively. Then, to obtain the rotational corrections of (say) the Love number, one needs to calculate the corresponding corrections to the mode eigenfrequencies and eigenfunctions (see \citealt{2024MNRAS.527.8409P} for a systematic analysis). Up to first order in the angular velocity, the Love number \eqref{Love number in phase space with rotation} is expressed as
\begin{equation}
	k_{lm}\approx\frac{2\pi G}{(2l+1)R^{2l+1}}\frac{e^{im[\Phi(t)-\Omega_s t]}}{v_{lm}}
    \sum_{n'} Q^{(0)}_{n'} \left[
        c^{(0)}_{n'_+} + (-1)^m c^{(0)*}_{n'_-}
        +c^{(1)}_{n'_+} + (-1)^m c^{(1)*}_{n'_-}
        +\frac{1}{Q^{(0)}_{n'}} \left( Q^{(1)}_{n'_+ l} c^{(0)}_{n'_+} + (-1)^m Q^{(1)}_{n'_- l} c^{(0)*}_{n'_-} \right)
    \right],
    \label{Love number in phase space slow rotation}
\end{equation}
with the superscripts denoting the rotational order of the corresponding quantity. The equations of motion for the zeroth- and first-order piece of the amplitude $c_n$ may be derived from \cref{phase amplitude equation of motion} and are given by
\begin{gather}
	\dot{c}^{(0)}_n-i\omega^{(0)}_n c^{(0)}_n = -\frac{i}{\mathcal{B}^{(0)}_n} v_{lm}(t) Q^{(0)}_n e^{-im[\Phi(t)-\Omega_s t]} 
	\label{mode amplitude zeroth order equation of motion} \\
	\shortintertext{and}
	\dot{c}^{(1)}_n-i\omega^{(0)}_n c^{(1)}_n = 
    i\omega^{(1)}_n c^{(0)}_n 
    -\frac{i}{\mathcal{B}^{(0)}_n} v_{lm}(t) Q^{(0)}_n \left( 
        \frac{Q^{(1)}_{nl}}{Q^{(0)}_n} 
        -\frac{\mathcal{B}^{(1)}_n}{\mathcal{B}^{(0)}_n}
    \right) e^{-im[\Phi(t)-\Omega_s t]}
    \label{mode amplitude first order equation of motion}
\end{gather}
respectively. \Cref{Love number in phase space slow rotation} can be simplified if we take a closer look at the formalism. First, note that we have already assumed that $Q^{(0)}_{n_+}=Q^{(0)}_{n_-}\equiv Q^{(0)}_n$; this should be true as long as the same normalisation is used for the modes (at zeroth order), namely $\mathcal{A}_{n_+}^{(0)2}=\mathcal{A}_{n_-}^{(0)2}$. Second, the first-order correction to the multipole moment can be decomposed as
\begin{equation}
	Q^{(1)}_{nl}=\frac{\mathcal{B}^{(1)}_n}{2\mathcal{B}^{(0)}_n} Q^{(0)}_n + \tilde{Q}^{(1)}_{nl} \ ,
\end{equation}
a consequence of the fact that the first-order correction to the mode eigenfunction, $\xi_n^{(1)i}$, has a component along $\xi_n^{(0)i}$ (which may be eliminated by normalising the first-order eigenfunctions such that $\mathcal{B}^{(1)}_n=0$, leading to further simplifications). Third, since first-order eigenfrequency and eigenfunction corrections scale with $m\Omega_s$, we must have $\tilde{Q}_{n_- l}^{(1)}=-\tilde{Q}_{n_+ l}^{(1)}\equiv -\tilde{Q}_{nl}^{(1)}$. Finally, the solution to the first-order equation of motion \eqref{mode amplitude first order equation of motion} can be written as
\begin{equation}
	c_n^{(1)}=\left(\frac{Q^{(1)}_{nl}}{Q^{(0)}_n}-\frac{\mathcal{B}^{(1)}_n}{\mathcal{B}^{(0)}_n}\right) c_n^{(0)} + \tilde{c}_n^{(1)} \ ,
\end{equation}
where $\tilde{c}_n^{(1)}$ is the solution to
\begin{equation}
	\dot{\tilde{c}}^{(1)}_n-i\omega^{(0)}_n \tilde{c}^{(1)}_n = i\omega^{(1)}_n c^{(0)}_n,
	\label{mode amplitude first order equation of motion modified}
\end{equation}
which is simply given by
\begin{equation}
	\tilde{c}_n^{(1)}=i\omega_n^{(1)} e^{i\omega_n^{(0)}t} \int_{t_0}^t c_n^{(0)} e^{-i\omega_n^{(0)}t'} dt'.
\end{equation}
Based on the above, \cref{Love number in phase space slow rotation} for the Love number is written as
\begin{equation}
	k_{lm}\approx\frac{2\pi G}{(2l+1)R^{2l+1}}\frac{e^{im[\Phi(t)-\Omega_s t]}}{v_{lm}}
    \sum_{n'} Q^{(0)}_{n'} \left[
        c^{(0)}_{n'_+} + (-1)^m c^{(0)*}_{n'_-}
        +\tilde{c}^{(1)}_{n'_+} + (-1)^m \tilde{c}^{(1)*}_{n'_-}
        +\frac{2 \tilde{Q}^{(1)}_{n'l}}{Q^{(0)}_{n'}} \left( c^{(0)}_{n'_+} - (-1)^m c^{(0)*}_{n'_-} \right)
    \right].
    \label{Love number in phase space slow rotation simplified}
\end{equation}

In the absence of rotation, \cref{Love number in phase space slow rotation simplified} may be reduced to
\begin{equation}
    k_{lm}=\frac{2\pi G}{(2l+1)R^{2l+1}}\frac{e^{im\Phi}}{v_{lm}}
        \sum_{n'}Q_{n'}\left[c_{n'_+}+(-1)^m c_{n'_-}^*\right]. \label{Love number in phase space}
\end{equation}
The relation between the phase space amplitudes $c_n$ and the amplitudes $a_n$ defined by the decomposition \eqref{mode sum} is generally non-trivial, but in the non-rotating case it yields
\begin{equation}
    c_n = \frac{1}{2}\left(-\frac{i}{\omega_n}\dot{a}_n+a_n\right).
\end{equation}
From this relation we can infer that
\begin{equation}
    c_{n_+}+(-1)^m c_{n_-}^*=a_{n_+}, \label{relation between configuration and phase space amplitudes without rotation}
\end{equation}
where we used the fact that $a_{n_-}=(-1)^m a_{n_+}^*$. Then, \cref{Love number in phase space} takes us back to \cref{klm general}. Accordingly, from \cref{modEn} we can express the mode energy in phase space (for no rotation) as
\begin{equation}
    E_n=2\omega_n^2\left|c_n\right|^2\mathcal{A}_n^2
\end{equation}
(the same result may also be obtained from \cref{mode energy in phase space with rotation}). The expressions above for the Love number and the energy, formulated in terms of the phase space amplitudes, may be used to compare the analytical relations derived in the current work to those obtained by \citet{2024PhRvD.110b4039Y}.

\bsp 
\label{lastpage}
\end{document}